\documentclass[journal]{IEEEtran}
\usepackage{cite}
\usepackage{amsmath, amssymb, physics, bm}
\usepackage{graphicx}
\usepackage{braket}
\usepackage{url}
\usepackage{hyperref}
\usepackage{xcolor}
\usepackage{authblk}
\usepackage{caption}
\usepackage[normalem]{ulem}
\usepackage{cancel}
\usepackage{float} 
\usepackage{subcaption}  
\ifCLASSINFOpdf
\else
\fi

\begin{document}
%
\title{
Atom--Field Non-Markovian Dynamics in Open and Dissipative Systems:
An Efficient Memory--Kernel Approach Linked to Dyadic Green’s Function and CEM Treatments
}
%
%
%

\author{Hyunwoo Choi,~\IEEEmembership{Student Member,~IEEE,}        
        Jisang Seo,~\IEEEmembership{Student Member,~IEEE,}
        Weng C. Chew,~\IEEEmembership{Life~Fellow,~IEEE,}
        and~Dong-Yeop~Na,~\IEEEmembership{Member,~IEEE}
\thanks{H. Choi, J. Seo and D.-Y. Na are with the Department
of Electrical Engineering, Pohang University of Science and Technology, Pohang,
Gyeongsangbuk-do, 37673 Republic of Korea e-mail: (dyna22@postech.ac.kr).}
\thanks{W. C. Chew is with the Elmore Family School of Electrical and Computer Engineering, Purdue University, West Lafayette, IN, 47907 USA.}
}

%
%

\markboth{Journal of \LaTeX\ Class Files,~Vol.~14, No.~8, August~2015}%
{Shell \MakeLowercase{\textit{et al.}}: Bare Demo of IEEEtran.cls for IEEE Journals}
%



\maketitle

\begin{abstract}
In this work, we present a numerical framework for modeling single-photon emission from a two-level system in open and dissipative systems beyond the Markovian approximation. The method can be readily integrated into standard computational electromagnetic (CEM) solvers such as finite-difference time-domain (FDTD) and finite element method (FEM). We numerically verify the completeness of boundary- and medium-assisted modes in the modified Langevin noise formalism by reconstructing the imaginary part of the dyadic Green’s function through modal expansion in three dimensions. This reconstruction enables a first-principles description of atom–field interaction via the multi-mode Jaynes–Cummings model in open and dissipative environments. Within the single-excitation manifold, we show that the memory kernel of a two-level system is determined by the imaginary part of the Green’s function, implying that radiative modes alone govern the relevant dynamics. The proposed framework thus provides a Green’s function–based approach for describing atomic population and single-photon dynamics, directly compatible with Maxwell solvers. We then present concrete strategies for implementing our method in both FDTD and FEM frameworks, demonstrating its practical applicability. We further verify numerical results for a lossy Lorentz–Drude-type mirror, including both the case of a TLS near a finite-sized metallic mirror and that of a TLS centered in a Fabry–Perot cavity. This work establishes a rigorous foundation for incorporating quantum emitter dynamics into computational electromagnetics, thereby extending classical solvers toward quantum light–matter interactions.
\end{abstract}


%
\IEEEpeerreviewmaketitle

\section{Introduction}

\IEEEPARstart{S}{ingle-photon} emission from two-level systems (TLS) in open, dissipative environments has attracted considerable attention in quantum electrodynamics (QED). Accounting for dissipation and material losses is essential, as they critically shape light–matter interactions in nanophotonic platforms~\cite{lodahl2015interfacing,arcari2014near}, plasmonic structures~\cite{archambault2010quantum,franke2022qnmlossy}, and integrated quantum devices~\cite{chang2018colloquium,Lee2024ProgrammableSiPhotonic4Qubit,Roth2025LorenzGauge,Sohn2024Integrated,Sohn2025microring}. However, in such open and dissipative systems, the effective Hamiltonian becomes non-Hermitian, leading to complex eigenvalues. This non-Hermitian nature complicates canonical quantization, as it requires preserving equal-time commutation relations between canonical variables~\cite{franke2022qnmlossy}. Conventional methods circumvent these complexities by utilizing the quantum master equation as the standard framework, achieving analytical tractability via Markovian memoryless bath and weak-coupling approximations~\cite{breuer2002theory,gardiner2004quantum}. However, when late-time interaction effects such as scattering from electrically large structures or out-coupled field distributions become significant, the Markovian and weak-coupling assumptions underlying the master equation approach break down~\cite{deVega2017dynamics}. Similarly, semi-classical treatments, which quantize the atom but not the electromagnetic field, can yield vanishing field expectations even when physical radiation is present~\cite{Scully1997quantum,Mandel1995optical}. Taken together, these limitations highlight that the most fundamental and accurate approach would be to track the time evolution of the full quantum state vector from first principles (i.e., second quantization)~\cite{Na2021cqNMD,Na2023quantumEMLossy,Na2020quantumInfoPreserving,Roth2024FullwaveQED}.

To this end, the modified Langevin noise (M-LN) formalism, rooted in the fluctuation–dissipation theorem, provides a rigorous framework for quantizing the electromagnetic field in dispersive and absorbing media by introducing boundary-assisted and medium-assisted (BA--MA) field modes~\cite{Drezet2017QuantizingPolaritons,Miano2025QuantumEmitterDispersiveDielectric,DiStefano2001ModeExpansion,Miano2025SpectralDensities}. This formalism ensures the preservation of canonical commutation relations while consistently accounting for both material / radiation loss and frequency dispersion.
Recently, this BA--MA field framework connected to a multi-mode Jaynes–Cummings (MMJC) model was introduced as a physically grounded modal representation for open and lossy EM systems coupled to TLS, based on the M-LN formalism~\cite{Choi2025nonMarkovian}. Although this method enables precise analysis and has demonstrated the ability to model phenomena previously beyond the reach of conventional quantization schemes, it remains computationally expensive.

For these reasons, rather than explicitly evaluating the entire set of BA–MA modes, we propose a computationally efficient framework enabling the equivalent calculation of the atomic population and the single-photon amplitude from a two-level system in open and dissipative electromagnetic environments. 
Specifically, when the Hilbert space is truncated to the single-excitation manifold, which is a widely adopted framework in quantum optics and light–matter interaction studies~\cite{Svidzinsky2015quantumVsSemiclassical,Schneider2015waveguideQED,Yin2022singlePhotonGiantMolecule,TerradasBrianso2024WaveguideQED}, the atomic population is governed by the imaginary part of the Green’s function, while the single-photon amplitude is characterized through equivalent current densities.
A key assumption in our derivation is that the BA--MA field modes form a complete set with real-valued eigenfrequencies, enabling the reconstruction of the dyadic Green’s function through a modal expansion in general open and dissipative environments. However, to the best of our knowledge, this reconstruction—particularly for the imaginary part of the dyadic Green’s function—has not been rigorously validated in fully three-dimensional settings. Rather than proceeding with this assumption left unverified, we aim in this work to rigorously establish it first, and only then build our subsequent analysis upon a solid foundation. To this end, we numerically verify the identity in two representative scenarios involving a dispersive and absorbing dielectric sphere: an open domain with a lossy metallic nanosphere near a point emitter, requiring both BA and MA modes, and a closed PEC cavity containing the same lossy object, where only MA modes contribute.

Subsequently, we consider the MMJC model~\cite{Jaynes1963Comparison,Scully1997quantum} constructed from these modes. In the single-excitation manifold, the time evolution of the total quantum state is fully described by the probability amplitude of the excited atom and the single-photon components in each BA--MA mode, and the corresponding equations of motion can be cast into a compact integro-differential form incorporating a memory kernel describing the atomic population dynamics. Notably, it is revealed that the memory kernel is fundamentally equivalent to the imaginary part of the dyadic Green’s function. This relation shows that, rather than explicitly computing all BA--MA modes, the atomic population and the radiated field defined by out-coupling efficiency of TLS can be equivalently obtained through the Green’s function, which in turn provides the natural connection to Maxwell’s equations in both time and frequency domains.
This approach successfully integrates quantum effects into classical computational electromagnetic frameworks such as the finite-element (FEM) or finite-difference time-domain (FDTD) scheme. As a result, it enables a numerically tractable analysis of non-Markovian quantum electrodynamics in open and dissipative electromagnetic environments. In the following, we describe the implementation procedure within these numerical schemes and present numerical examples for a lossy Lorentz--Drude-type mirror or cavity, comparing the results with the corresponding analytic solution.

The remainder of this paper is organized as follows.
Section II introduces the theoretical framework based on the M-LN formalism and establishes the connection between the dyadic Green’s function and the boundary- and medium-assisted (BA–MA) mode expansion.
Section III details the implementation of the proposed framework within finite-difference time-domain (FDTD) and finite-element method (FEM) solvers, emphasizing practical realization.
Section IV verifies, in fully three-dimensional settings, the completeness of the BA–MA modes by comparing Green’s-function reconstructions with direct numerical evaluations.
Section V presents numerical examples involving lossy Lorentz–Drude mirrors including Fabry–Perot configurations, demonstrating its validity and effectiveness.
Finally, Section VI concludes with a summary of the findings and future research directions.

\section{Formulation}

In this section, we formulate the theoretical framework underlying our approach. We begin with a brief introduction to the M-LN formalism, followed by an analysis of the connection between the Green’s function and the BA–MA modes, emphasizing the conditions for completeness of the modal basis. Finally, we establish an efficient formulation for modeling single-photon emission from a quantum emitter within the Jaynes–Cummings model.

For simplicity, only the positive-frequency part of the operator is considered throughout the entire derivation. We also adopt the $e^{-i\omega t}$ time convention and employ natural units ($\hbar = c = \epsilon_0 = \mu_0 = 1$), in which the wavelength remains unchanged while the frequency is normalized by the physical speed of light. Nonetheless, we retain them explicitly for SI consistency.

\subsection{Modified Langevin noise formalism}

The original Langevin noise formalism (O-LN), also known as the Green's function approach, was proposed to complement previous microscopic models~\cite{Huttner1992Quantization,Gruner1996QEDEvanescent,Dung2000QEDLocalized}. Specifically, the O-LN formalism describes the relationship between the ladder operators, which diagonalize the monochromatic Hamiltonian, and the electric field operator through the Langevin noise current operator as

\begin{align}
\hat{\mathbf{E}}^{(+)}(\mathbf{r},\omega)=i\omega\mu_0\int_{V_m}d\mathbf{r}'\overline{\mathbf{G}}(\mathbf{r},\mathbf{r}',\omega)\cdot\hat{\mathbf{J}}^{(+)}_N(\mathbf{r}',\omega),
\label{eq:O-LN}
\end{align}
where, 
\begin{align}
\hat{\mathbf{J}}^{(+)}_N(\mathbf{r},\omega)=\sqrt{\frac{\hbar\epsilon_0\omega^2}{\pi}\text{Im}[\epsilon_\omega(\mathbf{r})]}\hat{\mathbf{f}}(\mathbf{r},\omega).
\label{eq:Current-Operator}
\end{align}
Here, the total Hamiltonian is diagonalized in terms of the following dressed ladder operator as
\begin{align}
\hat{H} = \int^\infty_0d\omega\int_{V_m}d\mathbf{r}'\hbar\omega\hat{\mathbf{f}}^\dagger(\mathbf{r}',\omega)\cdot\hat{\mathbf{f}}(\mathbf{r}',\omega).
\end{align}

\begin{figure}
    \centering
    \begin{subfigure}[b]{0.8\linewidth}
        \centering
        \includegraphics[width=\textwidth]{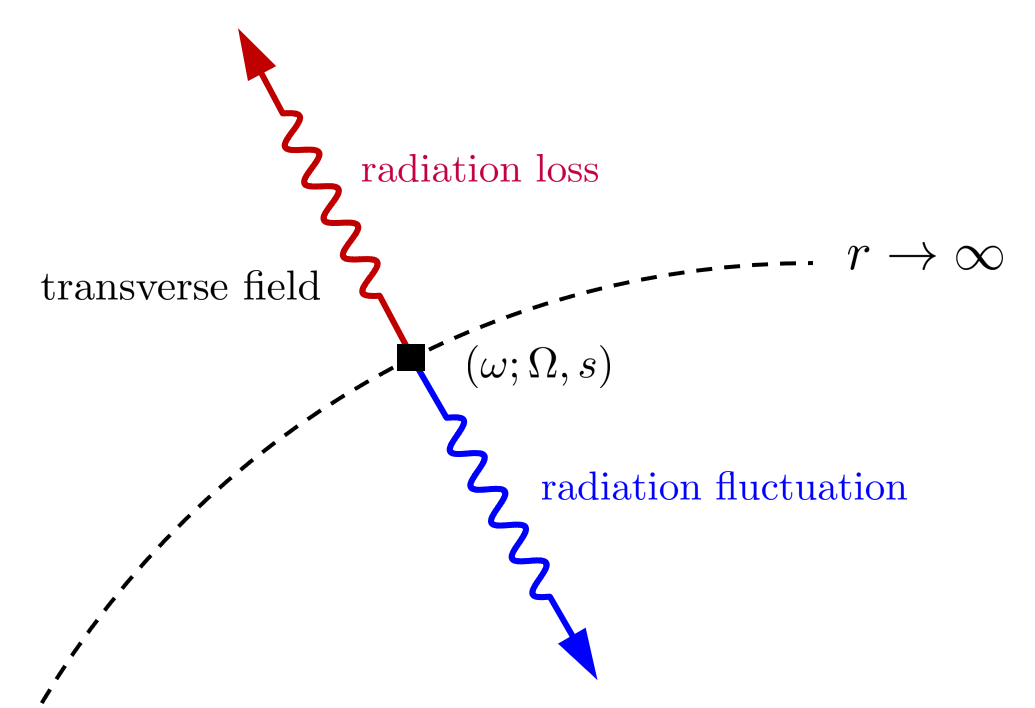}
        \caption{}
        \label{fig:open}
    \end{subfigure}
    \\
    \begin{subfigure}[b]{0.8\linewidth}
        \centering
        \includegraphics[width=\textwidth]{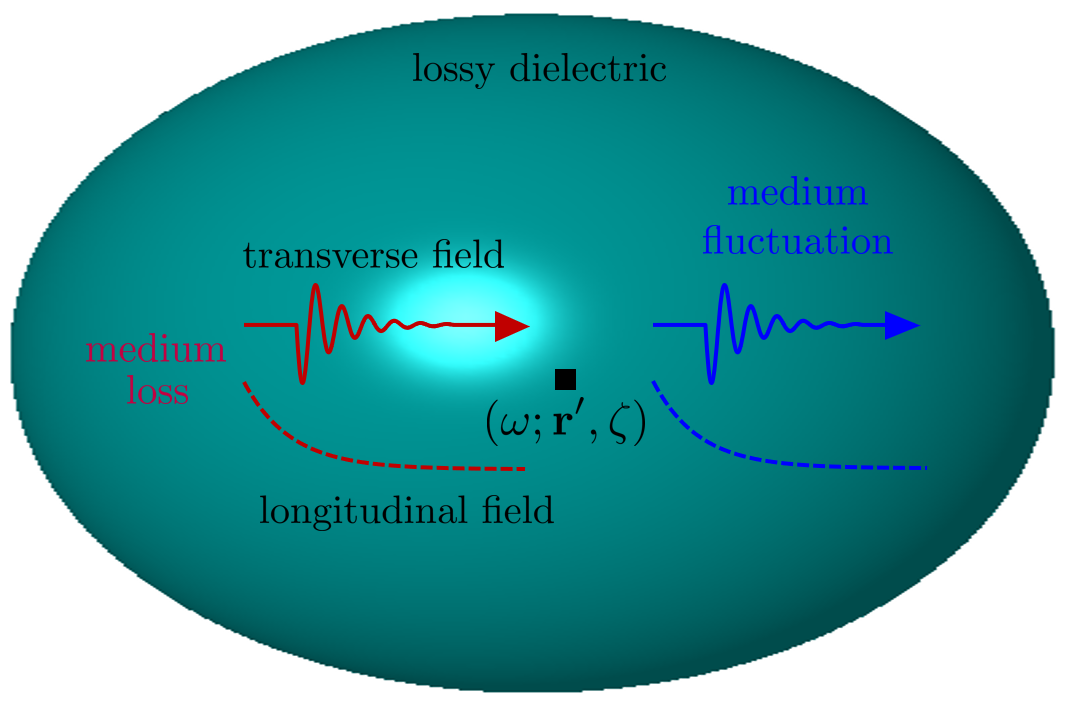}
        \caption{}
        \label{fig:open}
    \end{subfigure}
    \caption{Fluctuations based on the fluctuation--dissipation theorem: 
    (a) radiation-induced loss and corresponding field fluctuation, 
    and (b) medium-induced loss and associated material fluctuation.}    \label{fig:two_geometries}
\end{figure}

As shown in Eqs. (\ref{eq:O-LN}) and (\ref{eq:Current-Operator}), the strength of the O-LN formalism lies in its ability to establish a direct connection between the electric field operator and the ladder operator. This connection is mediated by the Green’s function and the noise current operator, forming a structure that mirrors the classical relation between the electric field and its current source. Therefore, one can physically interpret the field in this system as being determined by an infinite distribution of noise current sources, which originate from fluctuations associated with the medium’s losses.
However, as pointed out in previous work~\cite{Drezet2017QuantizingPolaritons,DiStefano2001ModeExpansion}, this relationship only holds in unbounded lossy dielectric media. When finite-sized lossy materials are introduced into the system, it becomes necessary to account for the effects of boundaries and openness—namely, the radiative degrees of freedom (DoFs) that carry energy away to infinity.
According to the M-LN formalism, these DoFs must be explicitly incorporated into the system via the introduction of BA-field operators. As a result, the total Hamiltonian should be modified to include the corresponding BA ladder operators, which are necessary for diagonalizing the full system as

\begin{align}
\hat{H} = \int_0^\infty d\omega \Bigg[ \int_{V_m} d\mathbf{r}'\, \hbar\omega\, 
\hat{\mathbf{f}}^\dagger(\mathbf{r}',\omega) \cdot \hat{\mathbf{f}}(\mathbf{r}',\omega) \nonumber \\
\quad + \oint_{S_k} d\mathbf{k}\, \hbar\omega\, 
\hat{\mathbf{a}}^\dagger(\mathbf{r}',\mathbf{k}) \cdot \hat{\mathbf{a}}(\mathbf{r}',\mathbf{k}) \Bigg].
\end{align}
Here, $\hat{\mathbf{a}}$ and $\hat{\mathbf{f}}$ denote the annihilation operators for the BA and MA fields, respectively. Fig.~\ref{fig:two_geometries} illustrates the origin of fluctuations arising from both radiative and medium-induced losses. These fluctuations consist of two distinct contributions, each of which fully satisfies the fluctuation–dissipation theorem. Accordingly, the electric field operator can be expressed as a linear combination of the MA and BA components.
\begin{flalign}
\hat{\mathbf{E}}^{(+)}(\mathbf{r},\omega)
=
\hat{\mathbf{E}}_{(\text{BA})}^{(+)}(\mathbf{r},\omega)
+
\hat{\mathbf{E}}_{(\text{MA})}^{(+)}(\mathbf{r},\omega),
\label{eqn:BA/MA_fields}
\end{flalign}
Rather than employing the Langevin noise operator formulation given in Eq.~(\ref{eq:O-LN}), we rewrite the operator by summing over all associated BA--MA modes and their corresponding ladder operators, in analogy with many existing quantization approaches~\cite{Scully1997quantum,Mandel1995optical,Garrison2008QuantumOptics}.

\begin{align}
\hat{\mathbf{E}}_{(\text{BA})}^{(+)}(\mathbf{r}, \omega)
&=
\int_{\mathbb{S}^2} k^2 d\Omega 
\sum_{s \in \{H, V\}} 
\mathbf{E}^{\text{(BA)}}_\omega(\mathbf{r}, \{\Omega, s\}) \nonumber \\
&\quad \times 
\hat{a}_\omega( \Omega, s), \label{eqn:BA_field} \\
\hat{\mathbf{E}}_{(\text{MA})}^{(+)}(\mathbf{r}, \omega)
&=
\int_{V_m} d\mathbf{r}'
\sum_{\xi \in \{x, y, z\}} 
\mathbf{E}^{\text{(MA)}}_\omega(\mathbf{r}, \{\mathbf{r}', \xi\}) \nonumber \\
&\quad \times 
\hat{f}_\omega(\mathbf{r}', \xi). \label{eqn:MA_field}
\end{align}
where $\mathbb{S}^2$ denotes the unit sphere surface, $\Omega$ the solid angle, $s$ the polarization index, $\mathbf{r}'$ the position of a noise point source, and $\xi$ its orientation. Here, degenerate modes ($\mathbf{E}_\omega$) refer to a set of distinct spatial modes ($\{\Omega,s\},\{\mathbf{r'},\xi\}$) that share the same eigenfrequency. At this stage, the problem becomes one of identifying the full set of BA and MA modes, each of which is defined as follows

\begin{align}
\mathbf{E}_\omega^{(\mathrm{BA})}(\mathbf{r},\{\Omega,s\})
&= \frac{i}{(2\pi)^{3/2}}
  \sqrt{\frac{\hbar\omega}{2}}\,
  \mathbf{E}_{\omega,\mathrm{tot}}(\mathbf{r},\{\Omega,s\}), \\[6pt]
\mathbf{E}^{(\mathrm{MA})}_\omega(\mathbf{r},\{\mathbf{r}',\xi\})
&= i\,k^2\,
  \sqrt{\frac{\hbar\,\chi_{I}(\mathbf{r}',\omega)}{\pi\epsilon_0}}\;
  \overline{\mathbf{G}}_{E}(\mathbf{r},\mathbf{r}',\omega)\cdot\hat{\xi}.
\end{align}
Each BA and MA degenerate mode can be obtained by solving the plane-wave scattering and point-source radiation problems, respectively — standard approaches in EM problems.

As a first step, we focus on the BA mode, which can be obtained by analyzing the plane-wave scattering problem. Specifically, the problem is to determine the total field modes $\mathbf{E}_{\text{tot}}$ produced by the scattering of plane waves impinging from specific inward radial directions $(\mathbf{k}_{\text{inc}})$ at infinity onto the material boundary. 
Let the incident field be defined as
\begin{flalign}
\mathbf{E}_{\omega,\text{inc}}(\mathbf{r},\{\Omega,s\}) 
= \hat{e}_s\, e^{i\mathbf{k}_{\text{inc}} \cdot \mathbf{r}},
\end{flalign}
Here, the direction of the incident field is given by,
\begin{align}
\mathbf{k}_{\omega,\text{inc}}(\Omega, s) 
&= 
\frac{\omega}{c} 
\Big( 
\sin\theta \cos\phi\, \hat{x}
+ \sin\theta \sin\phi\, \hat{y} + \cos\theta\, \hat{z} 
\Big). 
\label{eqn:inc_wavevector}
\end{align}
Accordingly, the scattered field due to the bounded material can be derived as follows
\begin{align}
&\nabla \times \nabla \times
  \mathbf{E}_{\omega,\text{sca}}(\mathbf{r}, \{\Omega, s\})
\nonumber \\[2pt]
&\quad - k^{2}\,\epsilon_{r}(\mathbf{r}, \omega)\,
  \mathbf{E}_{\omega,\text{sca}}(\mathbf{r},  \{\Omega, s\})
\nonumber \\[2pt]
&= i\,\omega \mu_{0}\,
  \mathbf{J}_{\text{BA}}(\mathbf{r}, \omega,\{\Omega, s\}),
\label{eqn:plane_wave_scattering_problem}
\end{align}
where the effective BA source is
\begin{flalign}
\mathbf{J}_{\text{BA}}(\mathbf{r},\omega,\{\Omega,s\}) 
= -i\omega \epsilon_0 \chi(\mathbf{r},\omega)\, \mathbf{E}_{\omega,\text{inc}}(\mathbf{r},\{\Omega,s\}).
\end{flalign}
Now, the total field is obtained by summing the incident and scattered fields as follows:
\begin{align}
\mathbf{E}_{\omega,\text{tot}}(\mathbf{r},  \{\Omega, s\}) 
&= 
\mathbf{E}_{\omega,\text{inc}}(\mathbf{r}, \{\Omega, s\}) 
 + 
\mathbf{E}_{\omega,\text{sca}}(\mathbf{r}, \{\Omega, s\}). 
\label{eqn:BA_total_field}
\end{align}

By contrast, each MA field mode is derived from the point-source radiation problem centered at a specific location ($r'$) within the lossy material, where the material's lossy properties $\chi_I$ are taken into account.
This can be obtained through 
\begin{align}
&\nabla \times \nabla \times
  \mathbf{E}^{(\text{MA})}_\omega,(\mathbf{r},  \{\mathbf{r}', \xi\})  
\nonumber \\[2pt]
&\quad - k^{2}\,\epsilon_{r}(\mathbf{r}, \omega)\,
  \mathbf{E}^{(\text{MA})}_\omega,(\mathbf{r},  \{\mathbf{r}', \xi\})
\nonumber \\[2pt]
&= i\,\omega \mu_{0}\,
  \mathbf{J}_{\text{MA}}(\mathbf{r}, \omega, \{\mathbf{r}', \xi\}).
\label{eqn:MA_mode_governing_eqn}
\end{align}
Where,
\begin{flalign}
\mathbf{J}_{\text{MA}}(\mathbf{r},\omega,\{\mathbf{r}',\xi\}) 
= \hat{\xi} \omega \epsilon_0 
\sqrt{\frac{\hbar \chi_I(\mathbf{r}',\omega)}{\pi \epsilon_0}}\, \delta(\mathbf{r} - \mathbf{r}').
\label{eqb:MA_source}
\end{flalign}

At this point, the total field operator can be expressed as given in Eqs.~(\ref{eqn:BA_field}) and (\ref{eqn:MA_field}). However, two important issues remain to be clarified in a three-dimensional system: (i) the systematic determination of all possible modes, and (ii) whether these BA--MA modes constitute a complete basis.
Regarding the first question, although relatively simple for obtaining individual modes, a key difficulty lies in the need to account for all possible directions $\mathbf{k}_{\text{inc}}$ in the case of BA modes, and all spatial points $\mathbf{r}'$ within lossy materials for MA modes, which form a continuum even for degenerate modes for a single eigenfrequency. As a possible approach, a coarse-graining strategy can be implemented using numerical methods such as the finite-element method (FEM). Nevertheless, due to the high computational cost, its application has so far been limited to one-dimensional systems~\cite{Na2020quantumInfoPreserving,Na2021cqNMD,Na2023quantumEMLossy,Choi2025nonMarkovian}. At the same time, the second question must be carefully addressed, as it directly concerns the fundamental assumption that the BA--MA ladder operators can completely diagonalize the total Hamiltonian. We verify this in the chapter IV; until then, we proceed under the assumption that the statement is valid.

\subsection{Green's function related to BA-MA mode expansion}
Modal expansion plays a central role in the analysis and design of electromagnetic systems across optical and radio-frequency (RF) regimes. In classical electrodynamics, however, performing mode decomposition in open and dissipative environments is non-trivial due to the appearance of complex-valued eigenfrequencies~\cite{Lalanne2018LightInteraction,Kristensen2020Modes,Vassallo1991OpticalWaveguide}. As a result, modal analysis is most commonly applied in closed, lossless systems such as cavities and waveguides, with workaround approaches like quasi-normal modes (QNMs) used for open systems~\cite{Ching1998QNMs, Wu2024DesigningResonators}.

In quantum electrodynamics, by contrast, modal expansion is indispensable since field quantization fundamentally relies on it. Importantly, the core concept of modal expansion remains the same in both classical analysis and quantum formulations. This observation implies that the M-LN formalism, although originally developed to describe dissipative QED, can also be viewed as a natural extension of modal expansion techniques to open and lossy systems, particularly in addressing issues associated with complex eigenfrequencies.
Deriving a complete set of BA–MA modes at real frequencies therefore provides a unified framework for accurately analyzing both classical and quantum electromagnetic responses. This construction is also physically significant, as it identifies the specific channels through which energy from a point-dipole source at $\mathbf{r}'$ is dissipated into radiative and material losses. With this foundation, we next examine how the BA–MA modal expansion connects to the dyadic Green’s function.

The frequency-domain dyadic Green’s function, $\overline{\mathbf{G}}(\mathbf{r}, \mathbf{r}', \omega)$, characterizes the electric field at position $\mathbf{r}$ in response to a point current source located at $\mathbf{r}'$ with orientation represented by the identity dyadic $\overline{\mathbf{I}}$, oscillating at $\omega$.
In linear, passive, and causal systems—such as open domains containing finite-sized absorbing or scattering structures—$\overline{\mathbf{G}}$ is analytic in the upper-half complex frequency plane and satisfies the Kramers–Kr\"{o}nig relations linking its real and imaginary parts.

The central ansatz of this work is a spectral representation of the imaginary part of the dyadic retarded Green's function in terms of the BA-MA field continuum modes:
\begin{flalign}
&
\frac{\mathcal{A}(\omega)}{\pi}
\operatorname{Im}\Bigl[\overline{\mathbf{G}}(\mathbf{r}, \mathbf{r}', \omega)\Bigr] 
=
\int_{\mathcal{D}_{\lambda}} d\lambda
\mathbf{E}^*_{\omega,\lambda}(\mathbf{r}) \otimes \mathbf{E}_{\omega,\lambda}(\mathbf{r}')
\label{eq:ImG_continuum}
\end{flalign}
where $\mathcal{A}(\omega)=\hbar \omega^2 \mu_0 $, $\lambda$ denotes the degeneracy index at $\omega$, which corresponds to one of degenerate BA-MA field modes, and $\mathcal{D}_{\lambda}$ is the degeneracy space. 
This ansatz holds if and only if the BA and MA fields constitute a complete continuum-mode basis for open and dissipative electromagnetic environments~\cite{sztranyovszky2025extending,chew2019greensdyadic}. When rewriting the above more specifically, one can have
\begin{flalign}
&
 \frac{\mathcal{A}(\omega)}{\pi}
\operatorname{Im}\Bigl[\overline{\mathbf{G}}(\mathbf{r}, \mathbf{r}', \omega)\Bigr] 
 \nonumber \\
&=
\sum_{s=1,2}
\int_{\mathbb{S}^2} d\Omega
k^2
\mathbf{E}_{\omega,\text{BA}}^{*}(\mathbf{r};\mathbf{k},s)
\otimes 
\mathbf{E}_{\omega,\text{BA}}(\mathbf{r}',\mathbf{k},s)
\nonumber \\
&+
\sum_{\zeta=x,y,z}
\int_{V_{m}} d\mathbf{r}''
\mathbf{E}^*_{\omega,\text{MA}}(\mathbf{r}; \mathbf{r}'',\zeta) 
\otimes 
\mathbf{E}_{\omega,\text{MA}}(\mathbf{r}''; \mathbf{r}',\zeta) 
\label{eqn:ansatz_ImG_BAMA}
\end{flalign}
These BA--MA modes are solutions to Maxwell's equations with outgoing wave or absorbing boundary conditions, and span the physical DoFs responsible for energy dissipation and radiation.

\subsection{BA--MA multimode Jaynes-Cummings model}
Consider a two-level system (TLS) with transition frequency \( \omega_a \), dipole moment \( \mathbf{d} \), and state \( \ket{e} \).
The TLS interacts with a set of quantized electromagnetic (EM) continuum BA--MA modes, denoted by $\mathbf{E}_{\omega,\lambda}$ with the degeneracy index $\lambda$.
The Hamiltonian under the rotating-wave approximation is given by
\begin{flalign}
\hat{H} 
&= 
\hbar \omega_a \hat{\sigma}_+ \hat{\sigma}_-
+ 
\int_{0}^{\infty}d\omega \int_{\mathcal{D}_{\lambda}}d\lambda 
\hbar \omega \hat{a}_{\omega,\lambda}^\dagger 
\hat{a}_{\omega,\lambda}
\nonumber \\
&+ 
\int_{0}^{\infty}d\omega \int_{\mathcal{D}_{\lambda}}d\lambda
\hbar g_{\omega,\lambda} \hat{\sigma}_{+} \hat{a}_{\omega,\lambda} + \text{H.c.} 
\end{flalign}
where the coupling constant is defined by
\begin{flalign}
g_{\omega,\lambda} 
= 
\frac{1}{i\hbar} 
\mathbf{d}_a 
\cdot 
\mathbf{E}_{\omega,\lambda}(\mathbf{r}_a).
\label{eqn:coupling_strength}
\end{flalign}
We assume the initial state is \( \ket{\psi(0)} = \ket{e, \{0\}} \) to model the TLS in the excited state initially, and can expand the total state in the truncated Hilbert space as
\begin{flalign}
\ket{\psi(t)} 
= 
C(t) \ket{e, \{0\}} 
+ 
\int_{0}^{\infty}d\omega \int_{\mathcal{D}_{\lambda}}d\lambda
D_{\omega,\lambda}(t) 
\ket{g, 1_{\omega,\lambda}}.
\end{flalign}
Within the rotating-wave approximation, the multimode Jaynes–Cummings (MMJC) model preserves the total excitation number throughout the entire time evolution.
From the Schrödinger equation, we obtain the following coupled equations of motion (EOMs) for the probability amplitudes $C(t)$ and $D_{\omega,\lambda}(t)$
\begin{flalign}
\dot{C}(t) 
=& 
- i \omega_a C(t) 
- i  \int_{0}^{\infty}d\omega \int_{\mathcal{D}_{\lambda}}d\lambda
g_{\omega,\lambda} D_{\omega,\lambda}(t), 
\label{eqn:SCE_coupled_eom1}
\\
&\dot{D}_{\omega,\lambda}(t) 
= 
- i \omega D_{\omega,\lambda}(t) 
- i g_{\omega,\lambda}^{*} C(t).
\label{eqn:SCE_coupled_eom2}
\end{flalign}
Solving the second equation formally, we obtain
\begin{flalign}
D_{\omega,\lambda}(t) = -i g_{\omega,\lambda}^* \int_0^t dt' e^{-i\omega(t - t')} C(t'). 
\label{eqn:C_gl_C_e_formal_sol}
\end{flalign}
Substituting \eqref{eqn:C_gl_C_e_formal_sol} back into \eqref{eqn:SCE_coupled_eom1}, the evolution of \( C(t) \) becomes
\begin{flalign}
\dot{C}(t) = 
-i \omega_a C(t) 
-i \int_0^t dt'
K_{\text{a}}(t-t')C(t')
\label{eq:Ct_dynamics}
\end{flalign}
where the memory kernel takes the form of
\begin{flalign}
K_{\text{a}}(t-t')
=
-i\frac{1}{\hbar^2}\int_{0}^{\infty} d\omega \int_{\mathcal{D}_{\lambda}}d\lambda
\left|\mathbf{E}_{\omega,\lambda}(\mathbf{r}_a)\cdot \mathbf{d}_a\right|^2
e^{-i\omega(t-t')}
\end{flalign}
with the use of the definition of $g_{\omega,\lambda}$ in \eqref{eqn:coupling_strength}.
Thanks to the identity \eqref{eq:ImG_continuum}, we can finally represent the memory kernel via
\begin{flalign}
K_{\text a}(t-t') &= -i\frac{\mu_0}{\pi\hbar}\int_{0}^{\infty} d\omega\,\omega^2\, e^{-i\omega(t-t')} 
\nonumber\\
&\quad\times\, \mathbf d_a \cdot \mathrm{Im}\!\left[\overline{\mathbf G}(\mathbf r_a,\mathbf r_a,\omega)\right]\cdot \mathbf d_a .
\label{eqn:memory_kernel}
\end{flalign}
The physical meaning of the above memory kernel become clear.
It represents the radiative component of the time-domain Green's function for a point source taken by itself.
When denoting 
\begin{flalign}
\mathcal{E}_{\text{rad}}(t) = \int_{0}^{t} dt' K_{\text{a}}(t-t') C(t')
\label{eqn:E_rad}
\end{flalign}
which represents the radiated fields by the TLS via the dipole current source excluding the reactive components, one can finally have
\begin{flalign}
\dot{C}(t) = -i \omega_a C(t) - i\mathcal{E}_{\text{rad}}(t)
\label{eqn:C_eom}
\end{flalign}
The central task here is to evaluate $\mathcal{E}_{\text{rad}}(t)$ efficiently.

On the other hand, for applications such as engineering the out-coupling efficiency of a quantum emitter in a given structure, it is often useful to consider the single-photon amplitude, which can be defined as
\begin{flalign}
\mathbf{E}_{\text{spa}}(\mathbf{r},t)
&=
\mel
{g,\left\{0\right\}}
{\hat{\mathbf{E}}^{(+)}(\mathbf{r},t)}
{\psi(t)}
\nonumber \\
&=
\int_{0}^{\infty}d\omega \int_{\mathcal{D}_{\lambda}}d\lambda
D_{\omega,\lambda}(t)\mathbf{E}_{\omega,\lambda}(\mathbf{r}).
\label{eqn:SPA_rep1}
\end{flalign}
Substituting \eqref{eqn:C_gl_C_e_formal_sol} into \eqref{eqn:SPA_rep1}, one can express the single-photon amplitude by
\begin{flalign}
\mathbf{E}_{\text{spa}}(\mathbf r,t)
&=\frac{1}{\hbar}
\int_{0}^{t} \! dt' \, C(t')
\int_{0}^{\infty} \! d\omega \, e^{-i\omega (t-t')}
\nonumber \\
&\quad \times
\int_{\mathcal{D}_{\lambda}} \! d\lambda \;
\left[
\mathbf{d}_a^\ast \cdot
\mathbf{E}^\ast_{\omega,\lambda}(\mathbf r_a)
\right]
\mathbf{E}_{\omega,\lambda}(\mathbf r).
\end{flalign}
And this can be rewritten in much more physically intuitive way
\begin{flalign}
\mathbf{E}_{\text{spa}}(\mathbf{r},t)
&=i\hbar
\int_{0}^{t}dt'
\mathbf{K}_{f}(\mathbf{r},\mathbf{r}_a,t-t')
C(t')
\label{eqn:SPA_kernel_exp}
\end{flalign}
where
\begin{flalign}
\mathbf{K}_{f}(\mathbf{r},\mathbf{r}_a,t-t')
&= -i\frac{\mu_0}{\pi\hbar}\int_{0}^{\infty} d\omega\,\omega^2\, e^{-i\omega(t-t')} 
\nonumber\\
&\quad\times\,  \mathrm{Im}\!\bigl[\overline{\mathbf G}(\mathbf{r},\mathbf r_a,\omega)\bigr]\cdot \mathbf d_a^{*}
\label{eqn:spa_rep2}
\end{flalign}
with the use of \eqref{eq:ImG_continuum} and \eqref{eqn:coupling_strength}.
The above vectorial memory kernel is now for electric fields, more specifically, single-photon amplitudes.

It is worth noting that the single-photon amplitude does not time evolve with eigenfrequencies of the bare eigenmodes since EM modes are coupled with the TLS such that resulting eigenfrequencies are dressed.
Instead, information of the dressed eigenfrequencies is now embedded in the probability amplitude $D_{\omega,\lambda}(t)$.

\subsection{Equivalent current formulation}

After defining the single-photon amplitude, we interpret it as the quantum analogue of the classical electric field. Within this framework, the two-level atom functions as a dipole current source, with its effect mediated by the Green’s function evaluated at the TLS position. Equivalently, the time evolution of the single-photon amplitude, as defined in \eqref{eqn:SPA_rep1}, can be understood as being driven by an effective point dipole current source expressed in terms of the atomic population.
Therefore, we further make this correspondence explicit by deriving the form of the equivalent current density $\mathbf{J}_{\text{TLS}}$, which encapsulates the radiation properties of the TLS and serves as the quantum source term in the field equations.

The Fourier transform of  \eqref{eqn:SPA_kernel_exp} can be written by
\begin{flalign}
\tilde{\mathbf{E}}'_{\text{spa}}(\mathbf r,\omega)
&=
\mu_0\omega^2
\tilde{C}(\omega)
\mathrm{Im}\,\overline{\mathbf{G}}
  (\mathbf r,\mathbf r_a;\omega)
\cdot \mathbf{d}_a^\ast.
\end{flalign}
The above equation can be reinterpreted as a radiation integral involving only the radiative component of the explicit current density prescribed by a point current source contributed by the TLS,
\begin{flalign}
\tilde{\mathbf{E}}'_{\text{spa}}(\mathbf r,\omega)
=
i \omega \mu_0
\int d^3\mathbf{r}' \;
\text{Im}\bigl[\overline{\mathbf{G}}_E(\mathbf r,\mathbf r';\omega)\bigr]
\cdot \mathbf{J}_{\text{TLS}}(\mathbf r',\omega),
\label{eq:GtoE}
\end{flalign}
where
\begin{flalign}
\mathbf{J}_{\mathrm{TLS}}(\mathbf r,\omega)
=
- i \omega \;
\mathbf{d}_a^\ast \, \tilde{C}(\omega) \;
\delta(\mathbf r - \mathbf{r}_a).
\label{eq:J_tls_final_append}
\end{flalign}
The corresponding expression in the time domain can be finally written as
\begin{flalign}
\mathbf{J}^{(+)}_{\mathrm{TLS}}(\mathbf r,t)
&=
\mathbf{d}_a^\ast \, \dot{C}(t) \;
\delta(\mathbf r - \mathbf{r}_a).
\label{eq:J_tls_time_append}
\end{flalign}
where $\dot{C}(t)$ denotes the time derivative of the single-photon amplitude $C(t)$.

This result is consistent with the conventional macroscopic viewpoint, where a TLS is treated as a dipole moment and the current is identified with the time derivative of the dipole polarization~\cite{Scheel2008MacroscopicQED}. Our formulation extends this picture by explicitly incorporating the single-photon out-coupling efficiency in the dissipative regime, derived from first principles within a M-LN formalism, in full agreement with established theory.

In short, observing \eqref{eqn:E_rad} and \eqref{eqn:SPA_kernel_exp}, it is evident that 
\begin{flalign}
\mathcal{E}_{\text{rad}}(t)
=
-\frac{1}{\hbar}\mathbf{d}_{a}\cdot \mathbf{E}_{\text{spa}}(\mathbf{r}_a,t).
\label{eq:d_dot_E}
\end{flalign}
This implies that, rather than solving for $C(t)$ and $D_{\omega,\lambda}(t)$ directly--that is, by explicitly including the continuum BA-MA modes--one can instead integrate \eqref{eqn:C_eom} numerically, using \eqref{eqn:E_rad} which may be obtained from full-wave electromagnetic simulations marching in the time domain.

\subsection{Decomposition into Markovian vacuum decay and non-Markovian scattering memory effect}
Since the dyadic Green's function can be separated into free-space and scattering contributions,
\begin{equation}
\overline{\mathbf{G}} = \overline{\mathbf{G}}_0 + \overline{\mathbf{G}}_{\text{scatt}},
\end{equation}
the memory kernel also decomposes as
\begin{equation}
K_a(t-t') = K_0(t-t') + K_{\text{scatt}}(t-t').
\end{equation}
Within the Wigner--Weisskopf approximation, the free-space part yields a local exponential decay rate,
\begin{equation}
\int_0^t dt' \, K_0(t-t') C(t') \;\approx\; \frac{\Gamma_0}{2} \, C(t),
\end{equation}
where $\Gamma_0$ is the spontaneous emission rate in vacuum.  
The scattering part remains a nonlocal convolution term:
\begin{equation}
\int_0^t dt' \, K_{\text{scatt}}(t-t') \, C(t').
\end{equation}
Finally, the equation of motion for $C(t)$ in \eqref{eqn:C_eom} can thus be expressed as
\begin{equation}
\dot{C}(t) = -i \omega_a C(t) - \frac{\Gamma_0}{2} C(t)
- i\mathcal{E}_{\text{scatt}}(t)
\label{eqn:population_dynaamics}
\end{equation}
where
\begin{equation}
\mathcal{E}_{\text{scatt}}(t)=
\int_0^t dt' \, K_{\text{scatt}}(t-t') \, C(t').
\end{equation}
This compactly shows the decomposition into a Markovian vacuum decay term and a non-Markovian scattering memory effect.
Utilizing Eqs. \eqref{eqn:SPA_kernel_exp} and \eqref{eqn:spa_rep2} and the equivalent current interpretation, the third term in expression \eqref{eqn:population_dynaamics} is associated with the scattered field from the equivalent current, $\mathbf{J}^{(+)}_{\mathrm{TLS}}$ in Eq. \eqref{eq:J_tls_time_append}. This interpretation is the crucial link for developing the CEM treatments of emission dynamics of TLS.

\section{Numerical Implementation via CEM Techniques}
The physical meaning of $\mathcal{E}_{\text{rad}}(t)$ in \eqref{eqn:E_rad} is the sum of the self-radiation field and the scattered fields from nearby objects.
Note that these fields exclude the self-reactive contribution, since the memory kernel in \eqref{eqn:memory_kernel} and \eqref{eqn:spa_rep2} must be described by the imaginary part of the dyadic Green's function.

We present two possible approaches for numerically evaluating $\mathcal{E}_{\text{rad}}(t)$:  
(i) FEM simulations for a point current source, from which the imaginary part of the dyadic Green's function can be extracted explicitly at each frequency;  
(ii) FDTD simulations with a point current source with the total-field/scattered-field (TFSF) formulation. Here, we consider the self-radiation fields separately using the Markovian description based on the adiabatic Wigner-Weisskopf model while the scattered fields are modeled via the FDTD-TFSF simulations \cite{zhou2024simulating}.

Note that although surface-integral equation (SIE) methods with the method of moments (MoM) are also applicable, we omit this approach here.

\subsection{FEM approach}

Since the emitter dynamics are fully determined by the full spectrum of imaginary part of the dyadic Green’s function,
it is not necessary to separately decompose the Markovian vacuum decay and non-Markovian scattering 
memory effects using FEM. Instead, the imaginary component $\mathrm{Im}\,\overline{\mathbf G}(\mathbf r_a,\mathbf r_a,\omega)$
can be directly computed in the frequency domain and subsequently used to evaluate the memory kernel.

Starting from the definition of the stationary memory kernel \(K_{\mathrm a}(\tau)\),
its Fourier transform is given by
\begin{align}
\tilde K_{\mathrm a}(\omega)
&= \int_{-\infty}^{\infty} d\tau\, e^{i\omega\tau}\,K_{\mathrm a}(\tau)\nonumber\\
&= \frac{\mu_0}{\pi\hbar}\int_{0}^{\infty} d\omega'\,\omega'^2\, 
\mathbf d_a\!\cdot\!\text{Im}\!\big[\overline{\mathbf G}(\mathbf r_a,\mathbf r_a,\omega')\big]\!\cdot\!\mathbf d_a \nonumber\\
&\quad\times \int_{-\infty}^{\infty} d\tau\, e^{i(\omega-\omega')\tau}.
\end{align}
Applying the orthogonality relation of exponential functions yields
\begin{align}
\tilde K_{\mathrm a}(\omega)
&= \frac{2\mu_0}{\hbar}\,\omega^{2}\,
\mathbf d_a\!\cdot\!\text{Im}\!\big[\overline{\mathbf G}(\mathbf r_a,\mathbf r_a,\omega)\big]\!\cdot\!\mathbf d_a.
\label{eq:Ka_def}
\end{align}
Therefore, once the imaginary part of the dyadic Green’s function is known,
the frequency-domain correlation spectrum \(\tilde{C}(\omega)\)—the Fourier transform of \(C(t)\)—can be directly obtained as
\begin{flalign}
\tilde{C}(\omega)
= \frac{C(t=0)}{-i\omega + i\omega_a + \tilde{K}_a(\omega)}.
\label{eq:C_omega}
\end{flalign}

To evaluate \(\text{Im}\,\overline{\mathbf G}(\mathbf r_a,\mathbf r_a,\omega)\) numerically, 
we solve the governing equation for the dyadic Green’s function,
\begin{flalign}
& \nabla\times \nabla \times \overline{\mathbf G}(\mathbf r,\mathbf r_a,\omega)
- \frac{\omega^2}{c^2} \epsilon(\mathbf r,\omega)\, \overline{\mathbf G}(\mathbf r,\mathbf r_a,\omega)
= \overline{\mathbf I}\, \delta(\mathbf r-\mathbf r_a),
\label{eq:dyadic_G_eq}
\end{flalign}
Multiplying both sides by a test function $\mathbf W_i(\mathbf r)$ and integrating over the computational domain $V$
leads to the weak form
\begin{align}
\begin{split}
& \int_V (\nabla\times\mathbf W_i^*)\!\cdot\!(\nabla\times\overline{\mathbf G})\, dV \\
& \quad - \frac{\omega^2}{c^2}\!\int_V \epsilon_r(\mathbf r,\omega)\, \mathbf W_i^*\!\cdot\!\overline{\mathbf G}\, dV
= \mathbf W_i^*(\mathbf r_a).
\end{split}
\label{eq:weak_form_G}
\end{align}

The dyadic Green’s function is expanded in terms of curl-conforming vector basis functions (Whitney 1-forms) as
\begin{equation}
\overline{\mathbf G}(\mathbf r,\mathbf r_a,\omega)
\simeq \sum_j \mathbf W_j(\mathbf r)\, \mathbf g_j(\mathbf r_a,\omega),
\end{equation}
and substituting this expansion into Eq.~\eqref{eq:weak_form_G} gives the discrete FEM matrix equation
\begin{equation}
\left(\mathbf S - \frac{\omega^2}{c^2}\,\mathbf M_\epsilon\right)\, \mathbf g = \mathbf f,
\label{eq:FEM_G_matrix}
\end{equation}
where
\begin{align}
[\mathbf S]_{ij} &= \int_V (\nabla\times\mathbf W_i^*)\!\cdot\!(\nabla\times\mathbf W_j)\, dV,\\[3pt]
[\mathbf M_\epsilon]_{ij} &= \int_V \epsilon_r(\mathbf r,\omega)\, \mathbf W_i^*\!\cdot\!\mathbf W_j\, dV,\\[3pt]
[\mathbf f]_i &= \mathbf W_i^*(\mathbf r_a).
\end{align}
The stiffness matrix $\mathbf S$ represents the curl–curl operator,
the mass matrix $\mathbf M_\epsilon$ describes the material response,
and $\mathbf f$ corresponds to the delta-function source at $\mathbf r_a$.
Once this linear system is solved, the complex-valued $\overline{\mathbf G}$ is obtained directly,
and its imaginary component can be extracted.
This numerically obtained imaginary part can then be substituted into Eq.~\eqref{eq:Ka_def}
to compute the memory kernel \(\tilde{K}_a(\omega)\),
and subsequently used in Eq.~\eqref{eq:C_omega} to evaluate \(C(t)\) using inverse Fourier transform from which the spatially dependent electric field \(\mathbf{E}_{\mathrm{spa}}(\mathbf r,t)\) can be determined.

\subsection{FDTD approach}
The FDTD approach is more practical and intuitive as it directly solves for the system's dynamics in the time domain.
However, FDTD cannot distinguish the real and imaginary part of the dyadic Green's function.
To isolate the real part of the dyadic Green's function only, one can apply the total-field/scattered-field formulation to the decomposition of Markovian and non-Markovian terms in \eqref{eq:Ct_dynamics}.
The core idea is that the TFSF formulation can filter out the real part of the dyadic Green's function and account for the scattering part while self-radiation effect is considered via spontaneous emission term obtained after using the Wigner-Weisskopf theory.

When running FDTD simulations with a point current source and observing the resulting fields at the location of the point current source, there should be three different contributions: (i) self-reactive fields, (ii) self-radiative fields, and (iii) scattered fields from objects.
We should discard the self-reactive fields, which contribute to the (vacuum) Lamb shifts.
One possible trick is to employ the TFSF formulation. An auxiliary FDTD, confined to the scattering neighborhood of the atom and run concurrently with the main FDTD, computes the emitter’s self-field from the TLS. This field is injected into the main domain across the TFSF surface via the surface-equivalence theorem, ensuring that the TLS location in the main simulation evolves only the environment-induced scattered field, composed solely of radiative components. When this scattered field returns to the TLS location, the atomic population is updated using the local scattered-field amplitude and phase; because the self-field has been subtracted at the TFSF interface, the non-propagating reactive component at the emitter is eliminated.

\begin{figure}[t]
    \centering
    \includegraphics[width=\linewidth]{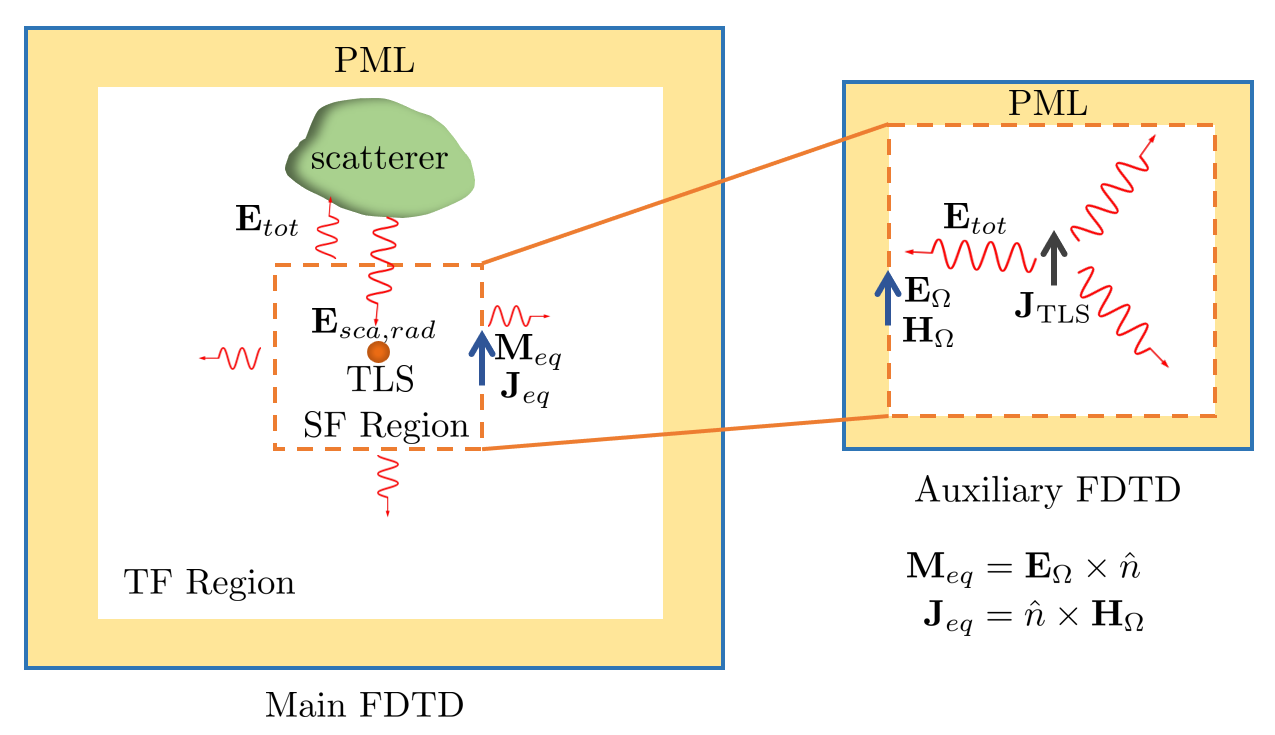}
    \caption{Schematic of the total-field/scattered-field (TFSF) formulation, where only the scattered field exists at the atom position.}
    \label{fig:TFSF}
\end{figure}

This strategy--referred to as FDTD for quantum emitters (FDTD-QE)--is not the first to update atomic populations via the scattered field using FDTD and TFSF, which has previously established a related scheme aimed at overcoming the limitations of the semi-classical viewpoint~\cite{zhou2024simulating}.
Inspired by that work, our derivation is grounded in a first-principles quantization framework that explicitly incorporates dissipative environments, thereby distinguishing it from the previous approach. By explicitly incorporating medium loss from MA modes, the Lorentz–Drude medium can be implemented in FDTD via an auxiliary differential equation for Lorentzian polarization. For BA modes, a perfectly matched layer can be introduced while preserving the integrity of the first-principles formulation. Fig.~\ref{fig:TFSF} provides an overview of the TFSF configuration.

In detail, a complex-valued FDTD scheme is employed to directly propagate the complex current density of the two-level system, \( \mathbf{J}^{(+)}_{\mathrm{TLS}} \), without separating its real and imaginary components in Eq.~\eqref{eq:J_tls_time_append}. The associated current density is explicitly evaluated from the central difference of the complex amplitude, yielding a temporally staggered quantity consistent with the Yee grid arrangement.
\begin{flalign}
\mathbf{J}_{\text{TLS}}^{n-\frac{1}{2}}
=
\mathbf{d}_a^\ast
\left(
\frac{C^n - C^{n-1}}{\Delta t}
\right)
\label{eq:J_discrete_update}
\end{flalign}
The atomic population dynamics, Eq.~\eqref{eqn:population_dynaamics} are correspondingly discretized as
\begin{flalign}
C^{n+1}
=
C^{n-1}
-
2\Delta t
\left(
i\omega_a C^{n}
+
\frac{\Gamma_0}{2}C^{n}
+
i\mathcal{E}_{\text{scatt}}^{n}
\right),
\label{eqn:C_update}
\end{flalign}
where the complex-valued amplitude $C^{n}$ represents the quantum state of the emitter at the $n$-th time step.
Based on the coupled dynamics defined by Eqs.~\eqref{eq:J_discrete_update} and~\eqref{eqn:C_update}, 
\(\mathbf{J}_{\text{TLS}}\) is assigned at the TLS position in the auxiliary FDTD. 
Before the scattered field generated by the structure reaches the emitter, 
the system evolves according to the analytic free-space emission dynamics, 
since \(\mathcal{E}_{\text{scatt}}^{n}=0\). 
Once the field \(\mathbf{E}_\text{SPA}\) from $\mathbf{J}_{\text{TLS}}$ reaches the boundary of the scattered-field domain \(\Omega\) in the auxiliary FDTD grid, 
the corresponding \(\mathbf{E}\) and \(\mathbf{H}\) components are transformed into equivalent surface currents following the TFSF formulation. 
This procedure ensures that only the scattered field exists within the SF region of the main FDTD grid, 
thereby providing the scattered electric field sampled at the emitter’s location. 
Using this field, the atomic population and the equivalent current are subsequently updated 
at the next time step in accordance with Eqs.~\eqref{eq:J_discrete_update} and~\eqref{eqn:C_update}.

To incorporate material dispersion and absorption into the same computational framework, 
the Lorentz--Drude (LD) medium is introduced within the auxiliary differential equation (ADE) for polarization. 
The total polarization $\mathbf{P}$ is expressed as 
\(\mathbf{P}=\mathbf{P}_{\mathrm D}+\mathbf{P}_{\mathrm L}\), 
where \(\mathbf{P}_{\mathrm D}\) and \(\mathbf{P}_{\mathrm L}\) represent the Drude (free-electron) 
and Lorentz (bound-electron) contributions, respectively. 
Each component evolves according to
\begin{align}
\frac{\partial^2 \mathbf{P}_{\mathrm D}}{\partial t^2}
+ \gamma_{\mathrm D}\frac{\partial \mathbf{P}_{\mathrm D}}{\partial t}
&= \epsilon_0\,\omega_{p,\mathrm D}^{2}\,\mathbf{E}, \\[3pt]
\frac{\partial^2 \mathbf{P}_{\mathrm L}}{\partial t^2}
+ \gamma_{\mathrm L}\frac{\partial \mathbf{P}_{\mathrm L}}{\partial t}
+ \omega_{0,\mathrm L}^{2}\,\mathbf{P}_{\mathrm L}
&= \epsilon_0\,\omega_{p,\mathrm L}^{2}\,\mathbf{E},
\end{align}
\begin{equation}
\mathbf D = \epsilon_0\,\epsilon_\infty \mathbf E + \mathbf P.
\end{equation}
where $\mathbf{D}$ is the electric displacement field, $\epsilon_\infty$ is the relative permittivity at infinite frequency, $\gamma_{\mathrm{D}}$ and $\gamma_{\mathrm{L}}$ are the damping constants accounting for material losses, $\omega_{p,\mathrm{D}}$ and $\omega_{p,\mathrm{L}}$ are the plasma frequencies related to the oscillator strengths, and $\omega_{0,\mathrm{L}}$ is the resonance frequency of the bound electrons.
When discretized in time, each polarization dynamics are expressed as
\begin{align}
\mathbf{P}_{\mathrm L}^{\,n+1}
&=
\frac{1}{1+\tfrac{\gamma_{\mathrm L}\Delta t}{2}}
\Biggl[
\bigl(2-\omega_{0,\mathrm L}^{2}\Delta t^{2}\bigr)\mathbf{P}_{\mathrm L}^{\,n}
\nonumber \\
&
-\bigl(1-\tfrac{\gamma_{\mathrm L}\Delta t}{2}\bigr)\mathbf{P}_{\mathrm L}^{\,n-1}
+\epsilon_0\,\omega_{p,\mathrm L}^{2}\,\Delta t^{2}\,\mathbf{E}^{\,n}
\Biggr]
\end{align}

\begin{align}
\mathbf{P}_{\mathrm D}^{\,n+1}
=
\frac{
2\,\mathbf{P}_{\mathrm D}^{\,n}
-\bigl(1-\tfrac{\gamma_{\mathrm D}\Delta t}{2}\bigr)\mathbf{P}_{\mathrm D}^{\,n-1}
+\epsilon_0\,\omega_{p,\mathrm D}^{2}\,\Delta t^{2}\,\mathbf{E}^{\,n}
}{
1+\tfrac{\gamma_{\mathrm D}\Delta t}{2}
}.
\end{align}
With the leap-frog staggering $\mathbf H^{n+\tfrac12}$ and $\mathbf E^n$, Ampère’s law is advanced in the auxiliary FDTD as
\begin{equation}
\mathbf D^{\,n+1}
=
\mathbf D^{\,n}
+
\Delta t\Bigl(\nabla\times \mathbf H^{\,n+\tfrac12}
-
\mathbf J_{\text{TLS}}^{\,n-\tfrac12}\Bigr),
\label{eq:Ampere_update}
\end{equation}
After updating $\mathbf P_{\mathrm D}^{\,n+1}$ and $\mathbf P_{\mathrm L}^{\,n+1}$ with the ADE recursions already given, the electric field is obtained algebraically from the constitutive law,
\begin{equation}
\mathbf E^{\,n+1}
=
\frac{
\mathbf D^{\,n+1}
-
\mathbf P_{\mathrm D}^{\,n+1}
-
\mathbf P_{\mathrm L}^{\,n+1}
}{
\epsilon_0\,\epsilon_\infty
}.
\label{eq:E_from_D}
\end{equation}
This approach keeps all dispersive and lossy effects within $\mathbf P$, while $\epsilon_\infty$ contributes only as an instantaneous scaling factor.

When the atomic transition period $T_a$ is much shorter than both the radiative lifetime $T_{\text{rad}}$ and the characteristic reinteraction time of the scattered fields $T_{\text{sca}}$, the envelope of $C(t)$ varies only slowly compared to the carrier oscillation. 
This situation corresponds to the weak-coupling regime, where the dynamics are approximately Markovian. 
It should be noted, however, that this case can also include delayed-feedback-induced non-Markovian effects, for example when the scattered field is reflected back from perfectly conducting (PEC) objects.
In this case, one can make the following approximation:
\begin{flalign}
\dot{C}(t) \approx -i\omega_a C(t).
\label{eqn:C_free_evol_approx}
\end{flalign}
In contrast to the complex-valued FDTD formulation presented earlier,  
this approximation enables the algorithm to be implemented entirely in real-valued form.  
In our original derivation, Eq.~\eqref{eq:J_tls_time_append} defines the equivalent current density in terms of the positive-frequency (analytic) field only,  
whereas the general FDTD formulation propagates real electromagnetic fields.  
Under this approximation, the driving current can therefore be expressed in real form by including the Hermitian-conjugate (H.c.) contribution.

\begin{align}
\mathbf J_{\mathrm{TLS}}(\mathbf r,t)
&= \mathbf J_{\mathrm{TLS}}^{(+)}(\mathbf r,t)
+ \mathbf J_{\mathrm{TLS}}^{(-)}(\mathbf r,t) \nonumber\\
&= \big( \mathbf d_a^\ast \, \dot C^\ast(t) + \mathbf d_a \, \dot C(t) \big)
\, \delta(\mathbf r-\mathbf r_a).
\label{eq:J_tls_with_hc_text}
\end{align}
Substituting \eqref{eqn:C_free_evol_approx} into \eqref{eq:J_tls_with_hc_text}, one can finally arrive at
\begin{align}
\mathbf J_{\mathrm{TLS}}(\mathbf r,t)
= 2\,\omega_a\,\mathrm{Im}\!\left[ C(t) \right]\;\mathbf d_a\;\delta(\mathbf r-\mathbf r_a).
\label{eq:J_tls_final}
\end{align}
With this approximation, the algorithm converges to an equivalent form of the FDTD–QE scheme~\cite{zhou2024simulating}, 
differing only by the incorporation of the ADE formulation for polarization.

Regardless of whether the fields are propagated in real or complex form, the algorithm consists of the standard FDTD scheme supplemented by two ADEs: (i) the Lorentz–Drude model for material polarization, and (ii) the atomic population dynamics. With both formulations, the complete algorithm proceeds through the following steps at each time iteration:

\begin{enumerate}
    \item Compute $C^{n+1}$ in the main FDTD from the scattered field $\mathcal{E}_{\text{scatt}}^{n}$ sampled at the TLS position. At the initial step ($\mathcal{E}_{\text{scatt}}^{n}=0$), the evolution follows the analytic free-space decay.
    
    \item Using the updated $C(t)$, evaluate $\mathbf{J}_{\mathrm{TLS}}^{n-\frac{1}{2}}$ at the TLS position in the auxiliary FDTD.
    
    \item Propagate the fields in the auxiliary FDTD using the ADE--FDTD algorithm, driven by $\mathbf{J}_{\mathrm{TLS}}^{n-\frac{1}{2}}$.
    
    \item Transfer the radiated field from the auxiliary FDTD to the main FDTD across $\Omega$ via equivalent surface currents.
    
    \item Propagate the fields in the main FDTD using the ADE--FDTD algorithm.
\end{enumerate}

\section{Numerical Verification of Completeness of BA--MA Modes in 3D}

To establish the validity of the proposed ansatz Eq.~\eqref{eq:ImG_continuum}, it is essential to verify that the BA--MA modes indeed form a complete basis in fully 3D settings. 
In most practical scenarios, however, analytic solutions to BA--MA field modes are not available, which necessitates numerical evaluation. 
To this end, we compute the modes by sampling over the relevant degeneracy space, and approximate the continuum expression on the right-hand side of \eqref{eqn:ansatz_ImG_BAMA} with a discrete summation. 
The resulting discrete form can then be directly compared against the numerically evaluated Green’s function.

\begin{align}
&
\frac{\mathcal{A}(\omega)}{\pi}\Im\bigl[\overline{\mathbf{G}}(\mathbf{r},\mathbf{r}',\omega)\bigr]
\nonumber \\
&
\approx
\sum_{s=1,2}
\sum_{i=1}^{N_\Omega}
\Delta\Omega
\frac{\omega^2}{c^2}
\mathbf{E}_{\omega,\mathrm{BA}}^{*}(\mathbf{r};\Omega_i,s)
\otimes
\mathbf{E}_{\omega,\mathrm{BA}}(\mathbf{r}';\Omega_i,s)
\nonumber
\\
&+
\sum_{\zeta=x,y,z}
\sum_{n=1}^{N_{m}}
\Delta V_{n}
\mathbf{E}_{\omega,\mathrm{MA}}^{*}(\mathbf{r};\mathbf{r}_n,\zeta)
\otimes
\mathbf{E}_{\omega,\mathrm{MA}}(\mathbf{r}';\mathbf{r}_n,\zeta)
\end{align}

\begin{figure}[t]
    \centering
    \begin{subfigure}[b]{.7\linewidth}
        \centering
        \includegraphics[width=\linewidth]{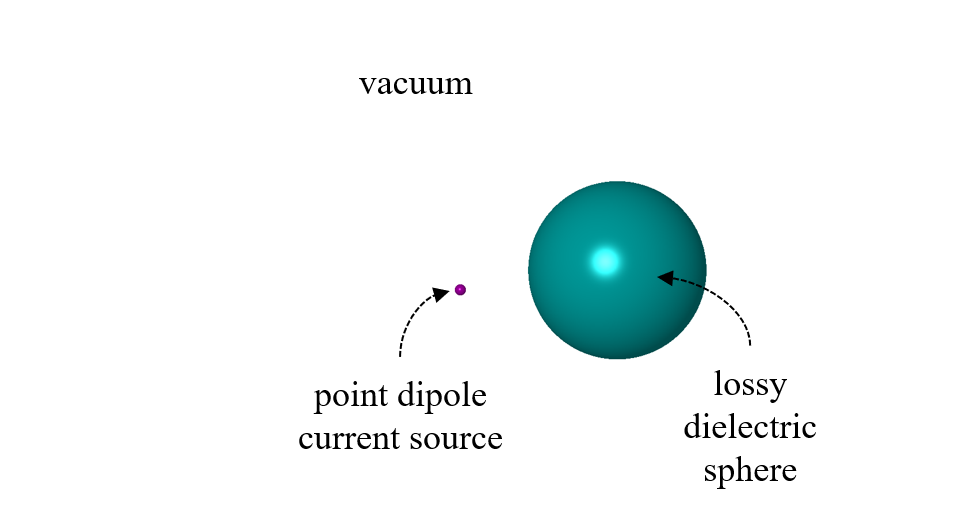}
        \caption{}
        \label{fig:scenario1}
    \end{subfigure}

    \begin{subfigure}[b]{.7\linewidth}
        \centering
        \includegraphics[width=\linewidth]{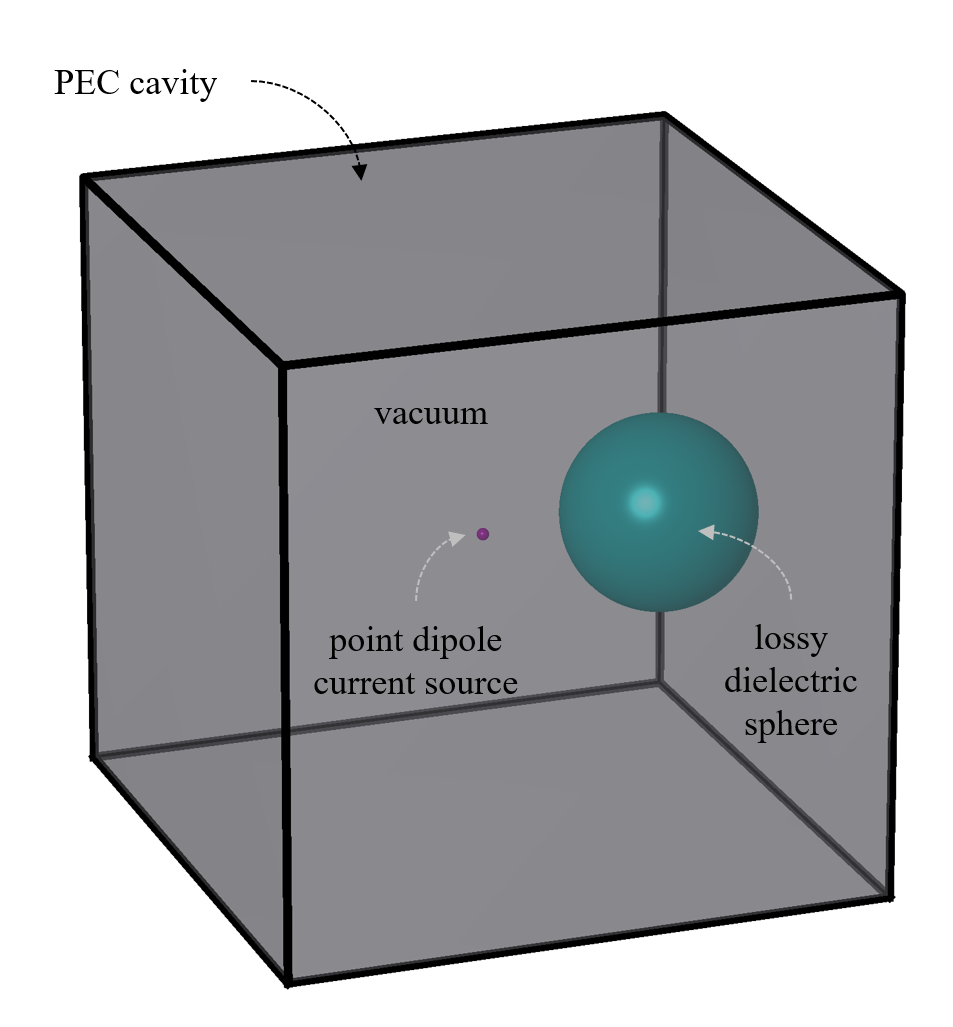}
        \caption{}
        \label{fig:scenario2}
    \end{subfigure}

    \caption{Representative 3D configurations used to verify the completeness of BA–MA modes: (a) a lossy dielectric sphere in free space, and (b) a lossy dielectric sphere inside a PEC cavity.}
    \label{fig:scenarios}
\end{figure}

The imaginary part of the dyadic Green's function can be numerically found by assuming a very short length edge based current source as a point current source and observe the radiating fields.
To rigorously establish the completeness of the BA–MA modal expansion, we examine two representative three-dimensional scenarios involving a point dipole source at position $\mathbf{r}_d$ near a lossy dielectric sphere. In the first case, the sphere is placed in free space, where both BA and MA modes are required since energy can dissipate through radiation as well as material absorption. In the second case, the sphere is enclosed within a vacuum-filled perfect electric conductor (PEC) cavity, so that only MA modes contribute, as radiative losses are fully suppressed.
Fig.~\ref{fig:scenarios} illustrates these two representative configurations considered in our numerical study.

\begin{figure}[t]
\centering            
\includegraphics[width=\linewidth]{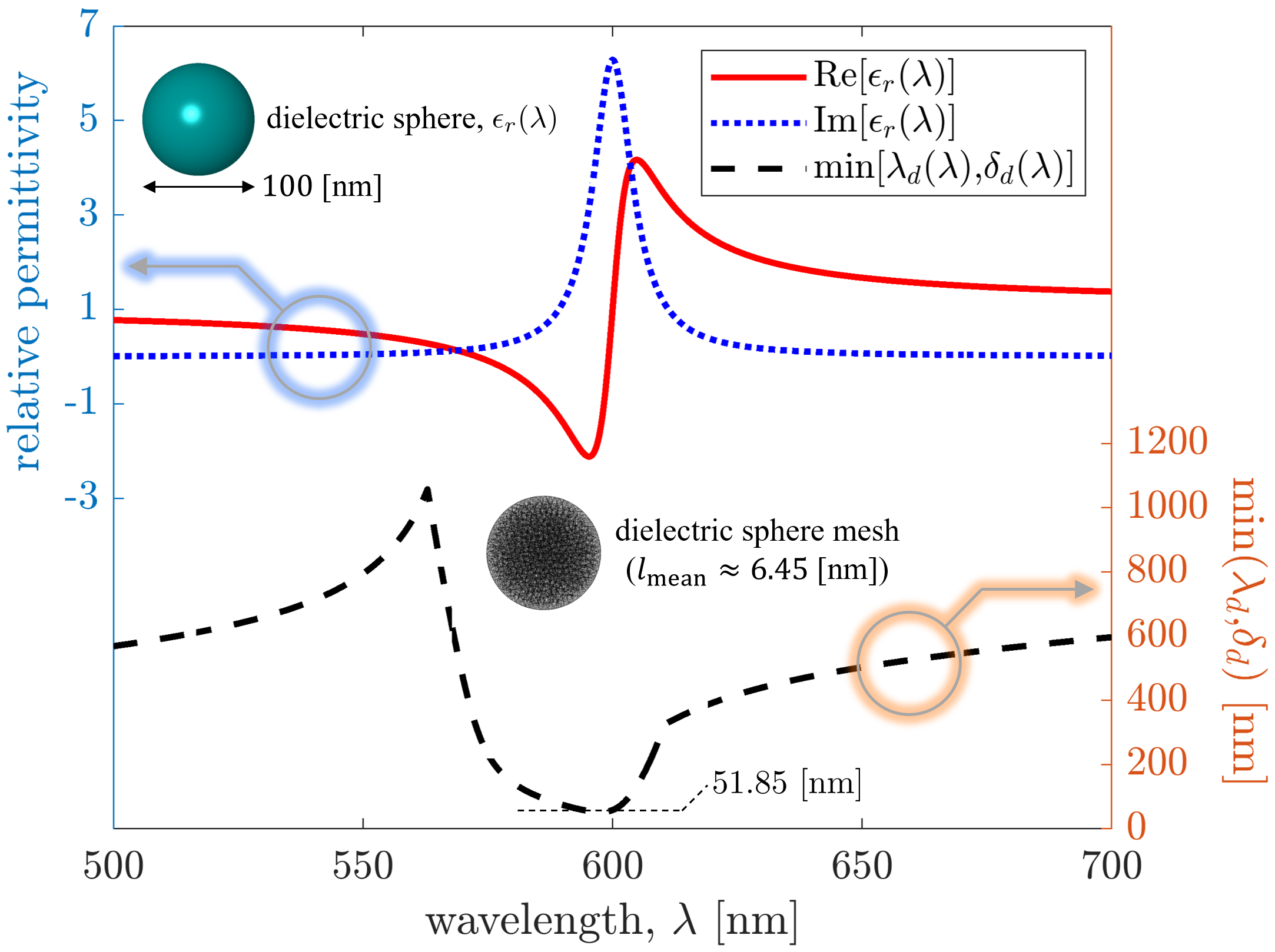}
\caption{Material properties of the dielectric sphere.}
\label{fig:open}    
\label{fig:material_property}
\end{figure}

To model a lossless dispersive dielectric, we consider $\text{SiO}_2$, which exhibits strong dispersion while remaining nearly lossless at optical wavelengths. The BA–MA field modes are computed numerically using the finite-element method (FEM), with the dielectric sphere discretized through an adaptive mesh refinement scheme to ensure accuracy. Specifically, the BA modes are obtained by solving the plane-wave scattering problem, while the MA modes are extracted from the radiation problem of a point current source embedded within the material.
The corresponding material dispersion properties are shown in Fig.~\ref{fig:material_property}.

In order to demonstrate the completeness of the BA–MA field modes, we evaluate the following discrete approximation of the ansatz in \eqref{eqn:ansatz_ImG_BAMA}
\begin{flalign}
&
\frac{\mathcal{A}(\omega)}{\pi}\Im\bigl[\overline{\mathbf{G}}(\mathbf{r},\mathbf{r}',\omega)\bigr]
\mathbf{d}
\cdot
\operatorname{Im}\Bigl[\overline{\mathbf{G}}(\mathbf{r}_d, \mathbf{r}_d; \omega)\Bigr] 
\cdot
\mathbf{d}
\nonumber \\
&
\approx
\sum_{s=1,2}
\sum_{i=1}^{N_\Omega} 
\Delta \Omega
\frac{\omega^2}{c^2}
\left|
\mathbf{d}
\cdot
\mathbf{E}_{\omega,\text{BA}}(\mathbf{r}_d;\Omega_i,s)
\right|^2
\nonumber \\
&+
\sum_{\zeta=x,y,z}
\sum_{n=1}^{N_{m}}
\Delta V_{n}
\left|
\mathbf{d}
\cdot
\mathbf{E}_{\omega,\text{MA}}(\mathbf{r}_d; \mathbf{r}_n,\zeta) 
\right|^2
\label{eqn:ansatz_ImG_BAMA_discrete}
\end{flalign}
Since 
\begin{flalign}
\mathbf{E}_{d\omega,\text{BA}}(\mathbf{r}_d;\Omega_i,s)
=
\mathbf{E}_{\omega,\text{inc}}
+
\int_{V_{m}}
\overline{G}(\mathbf{r}_d,\mathbf{r}')
\cdot
\mathbf{J}_{\text{BA}}(\mathbf{r}',\omega;\Omega_i,s)
\end{flalign}
For the BA contribution, the total field at the dipole location $\mathbf{r}_d$ is written as
\begin{flalign}
\mathbf{E}_{\omega,\text{BA}}(\mathbf{r}_d;\Omega_i,s)
=
\mathbf{E}_{\omega,\text{inc}}
+
\int_{V_{m}}
\overline{\mathbf{G}}^{T}(\mathbf{r}',\mathbf{r}_d)
\cdot
\mathbf{J}_{\text{BA}}(\mathbf{r}',\omega;\Omega_i,s)
\end{flalign}
By invoking the reciprocity relation of the dyadic Green’s function, the above expression can be equivalently rewritten. In practice, this means that the FEM solver needs to be run only once with a point current source placed at $\mathbf{r}_d$, while the resulting fields are recorded at $\mathbf{r}' \in V_m$ for use in the volume integral. In this formulation, the effective source current to be evaluated is given by
\begin{flalign}
\mathbf{J}_{\text{BA}}(\mathbf{r},\omega;\Omega_i,s)
=
\frac{\omega^2}{c^2}
\epsilon_r(\mathbf{r},\omega)
\mathbf{E}_{\omega,\text{inc}}(\mathbf{r};\Omega_i,s).
\end{flalign}
in the volume $V_{m}$ numerical quadrature.

Similarly, the MA field at the dipole location $\mathbf{r}_d$ due to a source at $\mathbf{r}n$ with polarization $\hat{\mathbf{e}}\zeta$ can be expressed as
\begin{flalign}
\mathbf{E}_{\omega,\text{MA}}(\mathbf{r}_d; \mathbf{r}_n,\zeta) 
=
\int_{V_{m}}
\overline{\mathbf{G}}(\mathbf{r}_d,\mathbf{r}')
\cdot
\mathbf{J}_{\text{MA}}(\mathbf{r}',\mathbf{r}_n,\zeta), 
\end{flalign}
Since 
\begin{flalign}
\mathbf{J}_{\text{MA}}(\mathbf{r},\mathbf{r}_n,\zeta)
=
\hat{\mathbf{e}}_{\zeta}
\sqrt{\frac{\operatorname{Im}\Bigl[\chi(\mathbf{r},\omega)\Bigr]}{\pi}}
\delta(\mathbf{r}-\mathbf{r}_n),
\end{flalign}

\begin{flalign}
\mathbf{E}_{\omega,\text{MA}}(\mathbf{r}_d; \mathbf{r}_n,\zeta) 
=
\overline{\mathbf{G}}(\mathbf{r}_d,\mathbf{r}_n)
\cdot
\hat{\mathbf{e}}_{\zeta}
\sqrt{\frac{\operatorname{Im}\Bigl[\chi(\mathbf{r}_n,\omega)\Bigr]}{\pi}}.
\end{flalign}
Using the reciprocal relation again,
\begin{flalign}
\mathbf{E}_{\omega,\text{MA}}(\mathbf{r}_d,; \mathbf{r}_n,\zeta) 
=
\overline{\mathbf{G}}^{T}(\mathbf{r}_n,\mathbf{r}_d)
\cdot
\hat{\mathbf{e}}_{\zeta}
\sqrt{\frac{\operatorname{Im}\Bigl[\chi(\mathbf{r}_n,\omega)\Bigr]}{\pi}}.
\end{flalign}
Thus, one can run a FEM solve for a point current source at $\mathbf{r}_d$ one time and record the radiation field at $\mathbf{r}_n\in V_{m}$ to compute the all the degenerate MA field modes at $\omega$.

\begin{figure}[t]
    \centering
    \begin{subfigure}[b]{\linewidth}
        \centering
        \includegraphics[width=\linewidth]{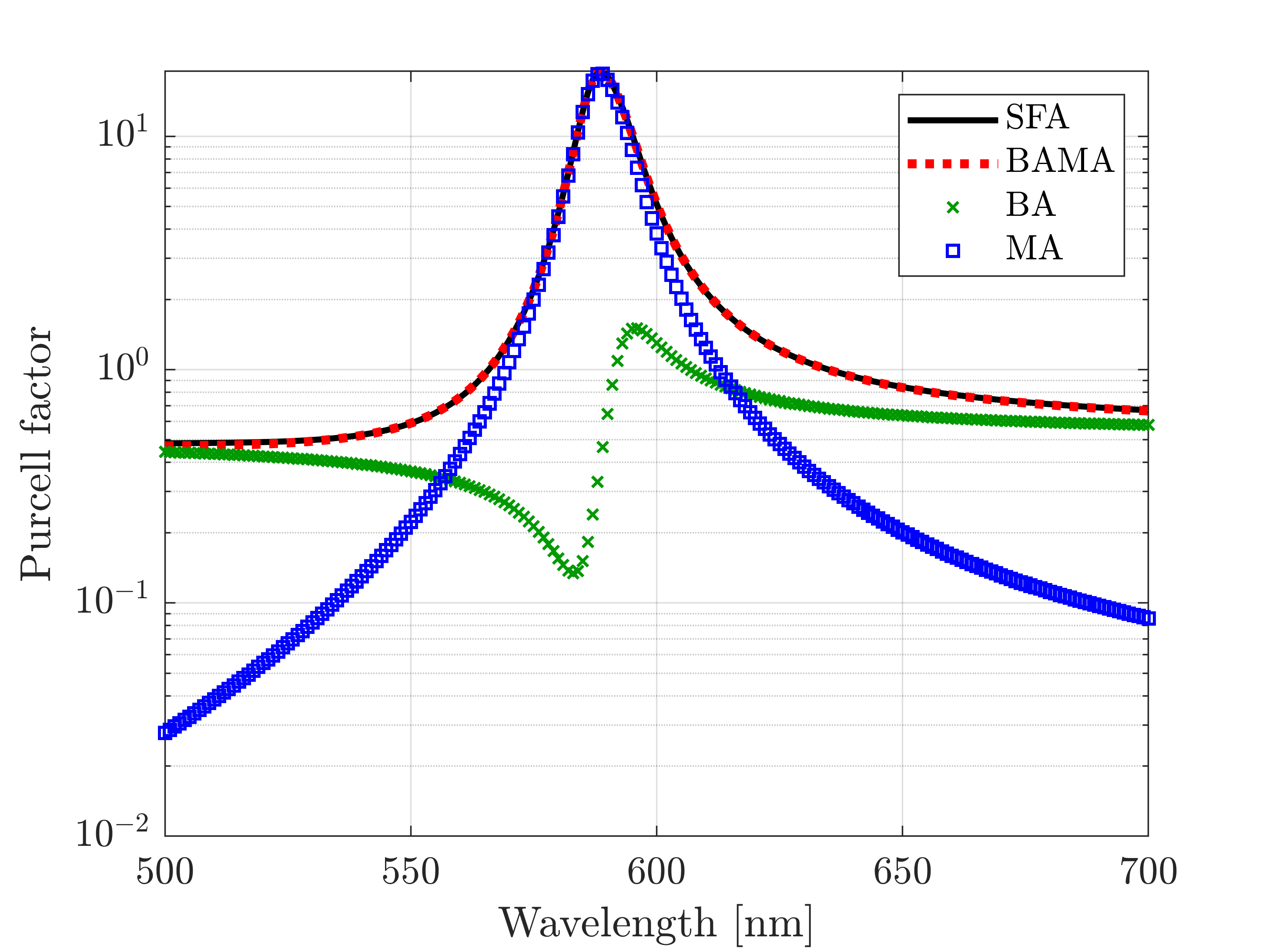}
        \caption{}
            \label{fig:open_lossy}
    \end{subfigure}

    \vspace{1em} 

    \begin{subfigure}[b]{\linewidth}
        \centering
        \includegraphics[width=\linewidth]{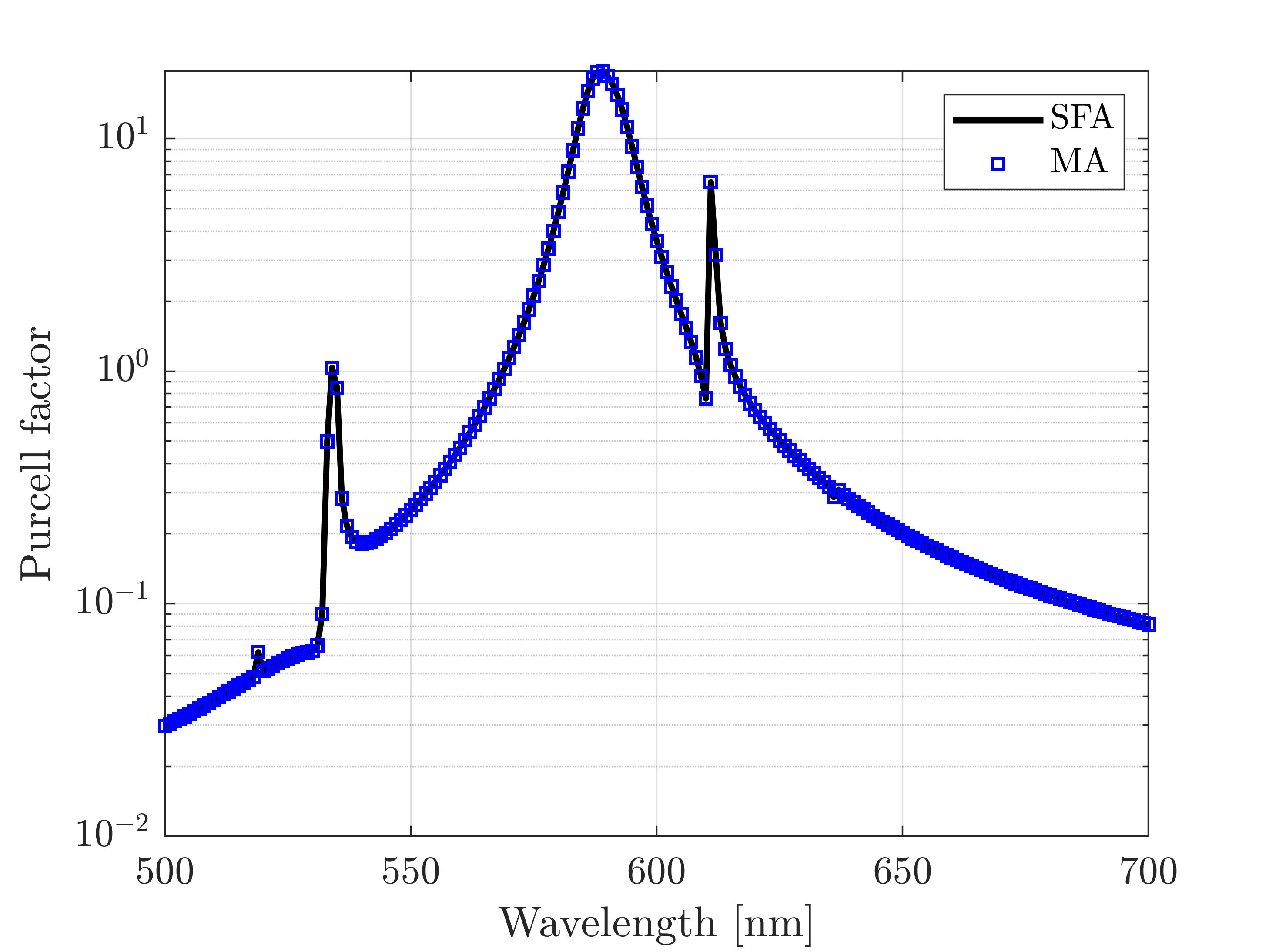}
        \caption{}
        \label{fig:PEC_lossy}
    \end{subfigure}

    \caption{Numerical verification of the BA--MA mode completeness: (a) a lossy dielectric in free space and (b) inside a PEC cavity.}    \label{fig:results}
\end{figure}
The reference Purcell factor can be evaluated using the spectral functional approach (SFA)~\cite{chew2019greensdyadic}, 
where the spectral function is defined as
\begin{align}
A(\mathbf{r}_a, \mathbf{r}_b, \omega) 
&= i \left( \mathbf{G}(\mathbf{r}_a, \mathbf{r}_b, \omega) 
       - \overline{\mathbf{G}}(\mathbf{r}_a, \mathbf{r}_b, \omega) \right) \notag \\
&= -2\,\mathrm{Im}\Bigl[ \mathbf{G}(\mathbf{r}_a, \mathbf{r}_b, \omega) \Bigr].
\end{align}
Based on the SFA, the spontaneous emission rate (SER) is given by
\begin{align}
\Gamma_{\mathrm{SFA}}(\omega_a)
= \frac{2\omega_a^{2}}{\hbar \epsilon_0 c^{2}}\,
\mathbf{d}\cdot\mathrm{Im}\!\Bigl[\mathbf{G}(\mathbf{r}_a,\mathbf{r}_a;\omega_a)\Bigr]\cdot\mathbf{d}^{*}.
\label{eq:SFA_rate}
\end{align}
Normalizing by the free-space rate,
\begin{equation}
\Gamma_0 = \frac{\omega_a^{3}|d|^{2}}{3\pi\epsilon_0 \hbar c^{3}},
\end{equation}
the Purcell factor becomes
\begin{align}
F_{\mathrm{P}}(\mathbf{r}_a,\omega_a;\hat{\mathbf{e}})
= \frac{6\pi}{k_0}\;
\hat{\mathbf{e}}\cdot\!\operatorname{Im}\Bigl[\overline{\mathbf{G}}(\mathbf{r}_a, \mathbf{r}_a; \omega_a)\Bigr] \cdot\hat{\mathbf{e}},
\end{align}
where $\hat{\mathbf{e}}$ denotes the dipole orientation and $k_0=\omega_a/c$.

Figs.~\ref{fig:open_lossy} and \ref{fig:PEC_lossy} show the FEM simulation results that verify the ansatz for the two representative scenarios illustrated in Fig.~\ref{fig:scenarios}: (a) a lossy dielectric sphere in free space and (b) the same sphere placed inside a PEC cavity. 
In both scenarios—the lossy dielectric sphere in free space and the same sphere enclosed within a PEC cavity—the Purcell factors obtained from the BA–MA modal expansion show excellent agreement with the reference values computed via the spectral functional approach. This consistent match across two fundamentally distinct physical situations (one allowing both radiative and absorptive channels, and the other suppressing radiative losses entirely) strongly validates the completeness of the proposed BA–MA framework and confirms the robustness of our numerical verification strategy.

\section{Simulation Results}
\begin{figure}[b]
\centering            
\includegraphics[width=\linewidth]{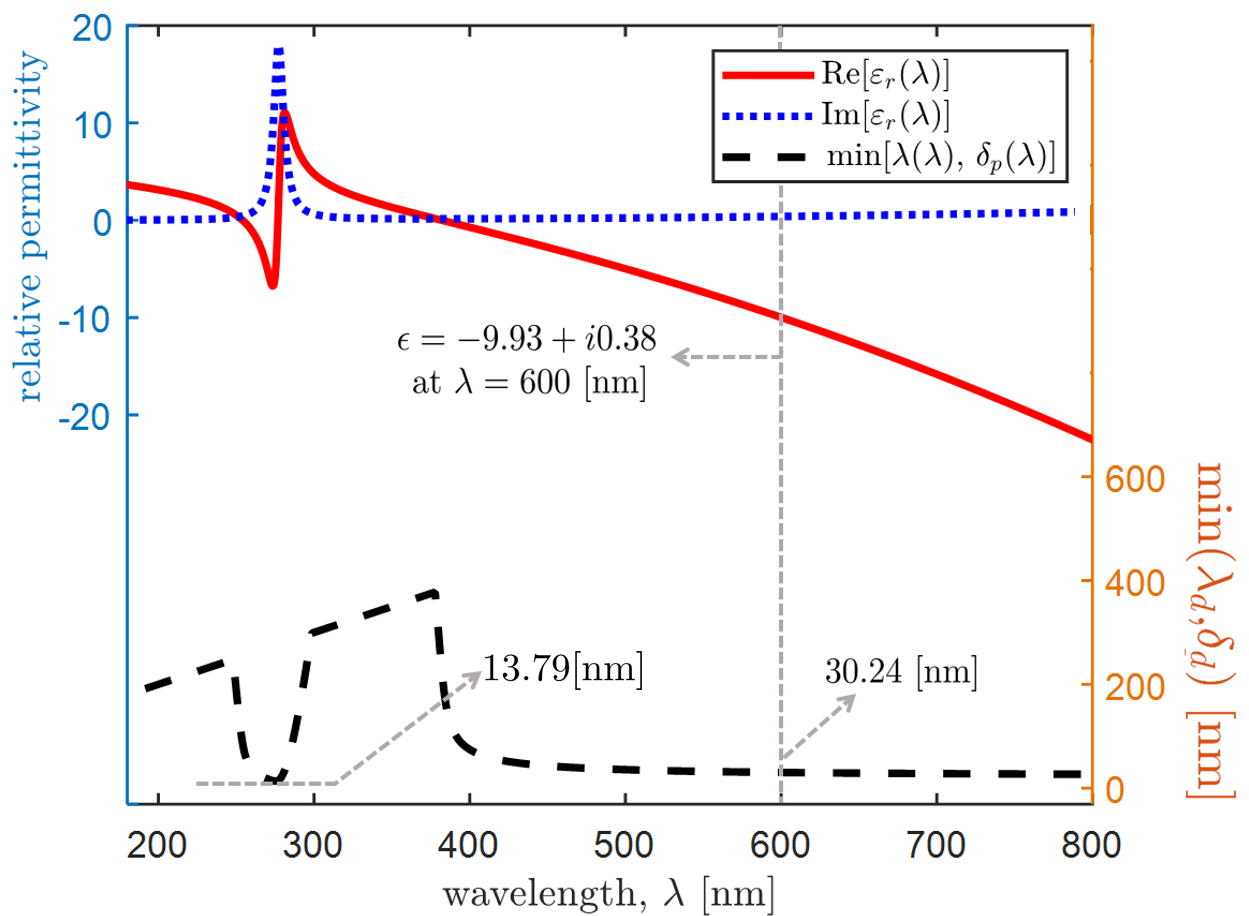}
\caption{Material properties of the metallic mirror.}
\label{fig:ld_epsilon}
\end{figure}
In this section, we present simulation results of spontaneous emission dynamics of a two-level system near or inside Lorentz--Drude-type lossy metallic mirrors. 
The relative permittivity of the metallic mirror is modeled by the Lorentz--Drude dispersion relation as
\begin{align}
\epsilon(\omega)
= \epsilon_\infty
- \frac{\omega_{p,\mathrm D}^2}{\omega^2 + i\,\gamma_{\mathrm D}\,\omega}
+ \frac{\omega_{p,\mathrm L}^2}{\omega_{0,\mathrm L}^2 - \omega^2 - i\,\gamma_{\mathrm L}\,\omega}.
\label{eq:LorentzDrude_epsilon}
\end{align}
where $\epsilon_\infty$ is the high-frequency background permittivity,
$\omega_{p,\mathrm D}$ and $\gamma_{\mathrm D}$ denote the plasma frequency and damping rate of the Drude (free-electron) contribution,
and $\omega_{p,\mathrm L}$, $\omega_{0,\mathrm L}$, and $\gamma_{\mathrm L}$ represent the effective plasma frequency, resonance frequency, and damping rate of the Lorentz (bound-electron) oscillator, respectively.
\subsection{TLS near finite-sized metallic mirror}
We first investigate the two-level system (TLS) placed near a finite-sized lossy mirror using FDTD–QE methods. 
The simulation domain of $1.2~\mu\mathrm{m} \times 1.2~\mu\mathrm{m} \times 3.6~\mu\mathrm{m}$ was discretized with a $6.67~\mathrm{nm}$ Yee cell as the spatial grid and enclosed by a PML.
The mirror material followed the Lorentz–Drude model with 
$\epsilon_\infty = 5.485$, 
$\omega_{p,\mathrm D} = 4.20\times10^{7}$,
$\gamma_{\mathrm D} = 2.43\times10^{5}$,
$\omega_{p,\mathrm L} = 1.61\times10^{7}$,
$\omega_{0,\mathrm L} = 2.27\times10^{7}$,
and $\gamma_{\mathrm L} = 6.33\times10^{5}~\mathrm{rad/s}$. 
As mentioned earlier, these frequency-related parameters are normalized by the physical speed of light compared to their physical values, since the simulation employs natural units where $c = 1$.
The TLS has a transition dipole moment $d_{a} = 5\times10^{-8}~\mathrm{C{\cdot}m}$, 
oriented along the $z$-axis, such that the dipole is polarized perpendicular to the mirror plane.
At this wavelength, the dielectric constant of the mirror material is $\epsilon(600~\mathrm{nm}) = -9.93 + i0.38$.
The mirror was modeled as a finite lossy metallic slab 
with a thickness $w$ of $150~\mathrm{nm}$ 
and lateral dimensions $l$ of $0.9~\mathrm{\mu m} \times 0.9~\mathrm{\mu m}$ 
in the $yz$ plane. 
The material dispersion characteristics for mirror are shown in Fig.~\ref{fig:ld_epsilon}.

\begin{figure}[t]
    \centering
    \includegraphics[width=\linewidth]{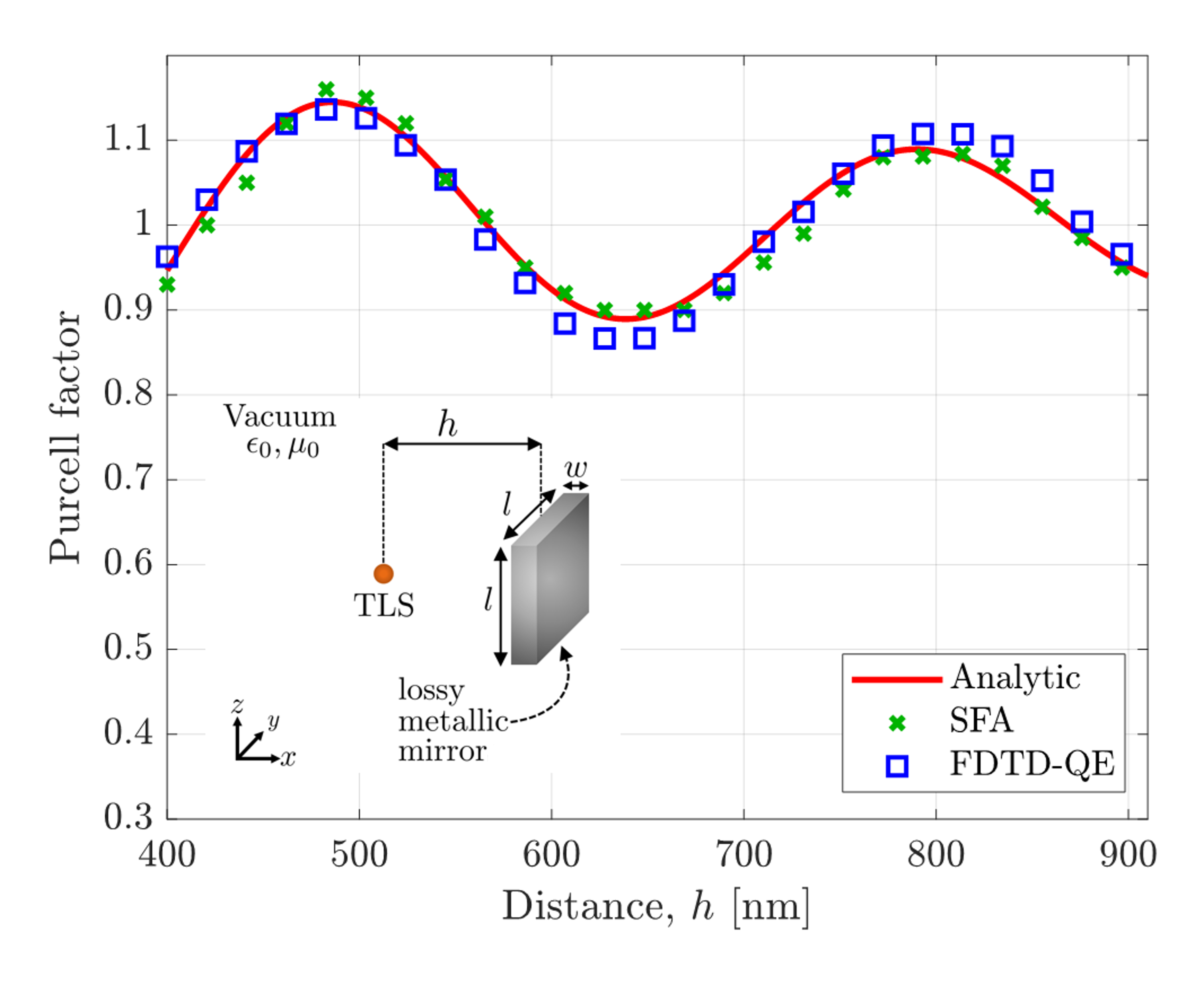}
    \caption{Purcell factor as a function of the emitter–surface distance $h$ for a metallic mirror geometry, considering both half-space and finite-sized configurations.}
    \label{fig:purcell_distance}
\end{figure}
\begin{figure}[!t]
    \centering
    \subfloat[]{%
        \includegraphics[width=0.48\linewidth]{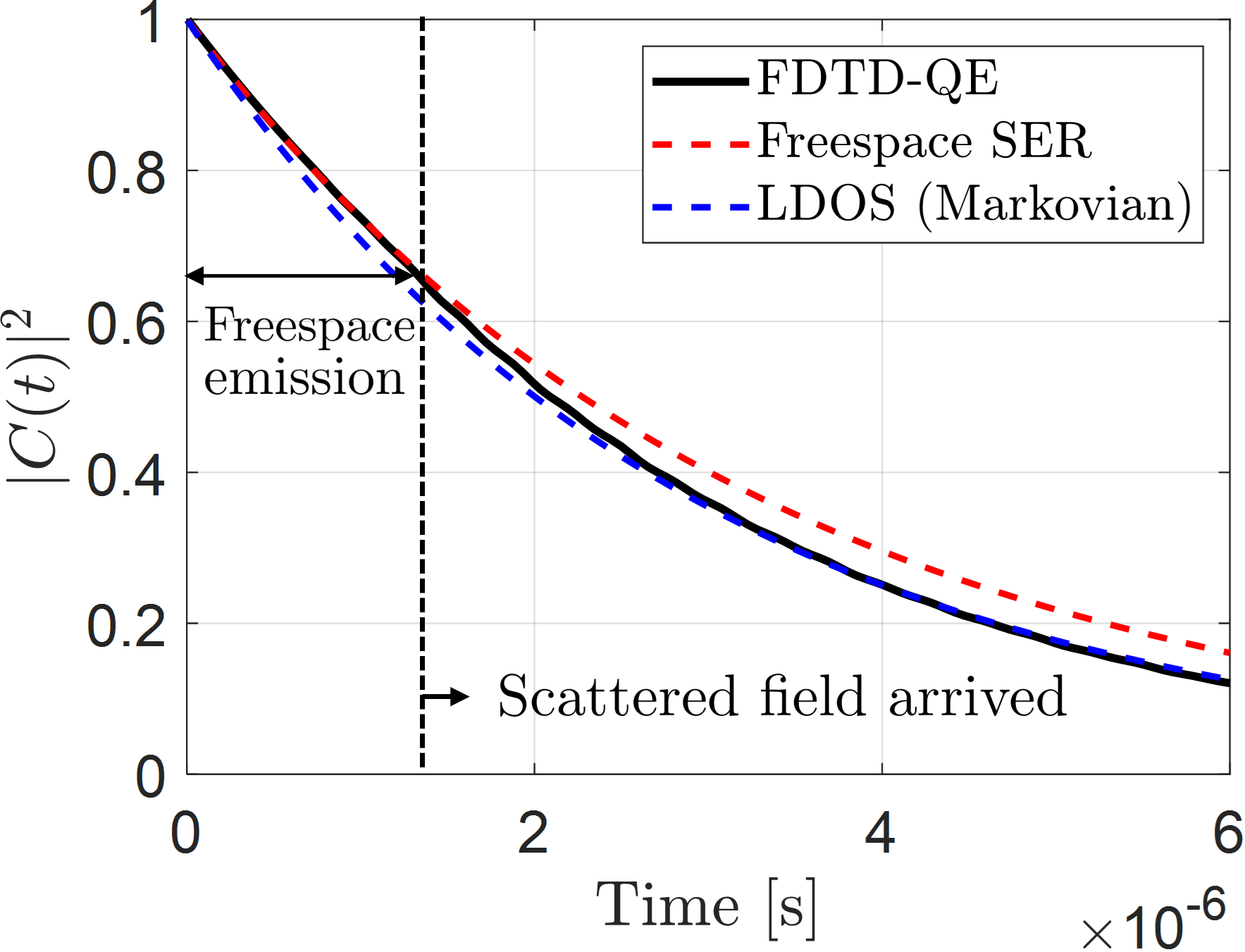}%
        \label{fig:pop_fast} 
    }
    \hfill 
    \subfloat[]{%
        \includegraphics[width=0.48\linewidth]{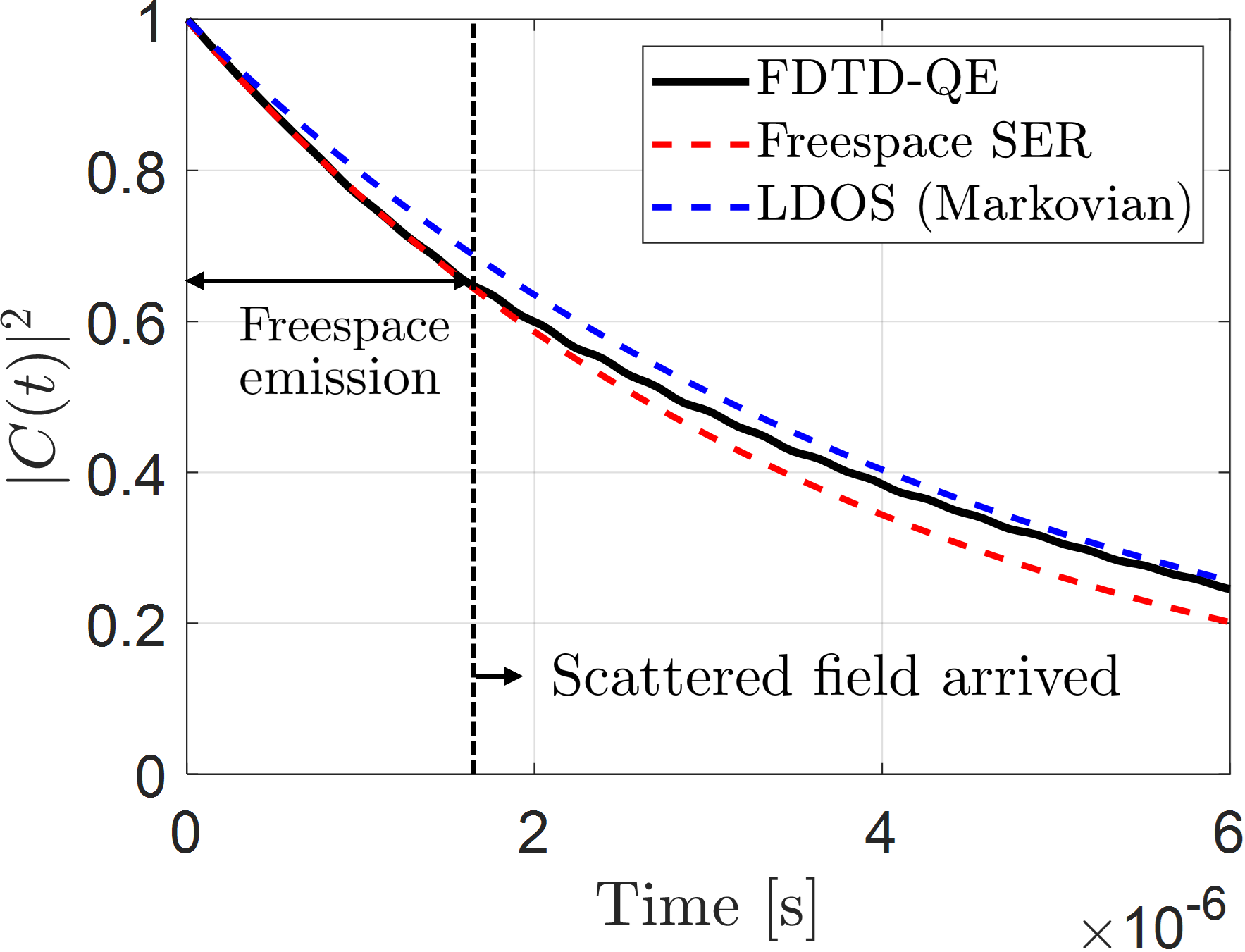}%
        \label{fig:pop_slow}
    }\\[1ex] 

    \subfloat[]{%
        \includegraphics[width=0.48\linewidth]{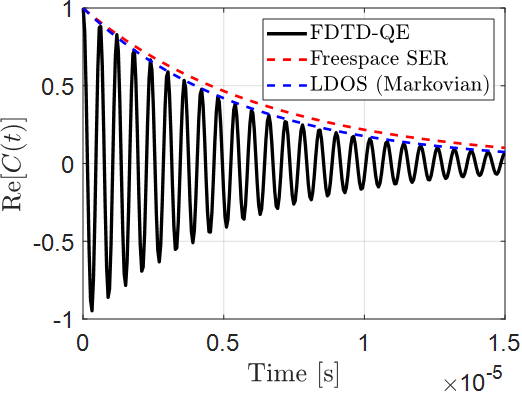}%
        \label{fig:ct_fast}
    }
    \hfill
    \subfloat[]{%
        \includegraphics[width=0.48\linewidth]{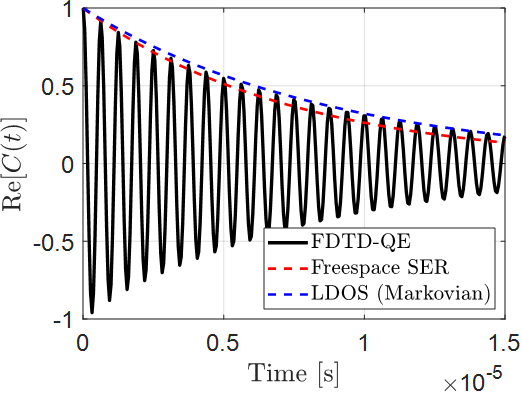}%
        \label{fig:ct_slow}
    }

    \caption{Time dynamics at emitter–surface distances giving the maximum ($h=482$ nm) and minimum ($h=627$ nm) Purcell factors: (a,c) maximum case and (b,d) minimum case. Panels (a,b) show the atomic population decay, while (c,d) show $\text{Re}[C(t)]$.}
    \label{fig:dynamics_all}
\end{figure}

Fig.~\ref{fig:purcell_distance} compares the Purcell factors obtained from the analytical Lorentz--Drude model, 
the SFA method based on Eq.~(\ref{eq:SFA_rate}), and the full FDTD--QE simulations.  
The analytical solution assumes an infinite planar metal interface 
following the formalism of Ref.~\cite{Jones2018Modified}.
The FDTD-QE simulation results deviate from a pure exponential decay due to non-Markovian effects. Consequently, we calculated the Purcell factor by determining the decay rate, $\Gamma = 1/\tau$, where $\tau$ is defined as the time at which the atomic population decays to $e^{-1}$.
All three results show consistent oscillatory behavior of the Purcell factor as a function of the emitter–mirror distance $h$, 
demonstrating reasonably good agreement between the numerical FDTD–QE simulations and the analytic model.
We also investigate the atomic population dynamics at two distances: the point of maximum Purcell factor (482 nm) and the point of minimum Purcell factor (627 nm). These results are detailed in Figs.\ref{fig:pop_fast} - \ref{fig:ct_slow}.
Unlike the local density of states (LDOS) method, which inherently assumes purely exponential atomic population dynamics, the FDTD-QE results show initial free-space emission before the arrival of the scattered field. Following the scattered field's arrival, the dynamics then converge to a behavior close to that predicted by the Markovian LDOS method. 
\begin{figure}[t!]
    \centering
    \begin{subfigure}[b]{\linewidth}
        \centering
        \includegraphics[width=\linewidth]{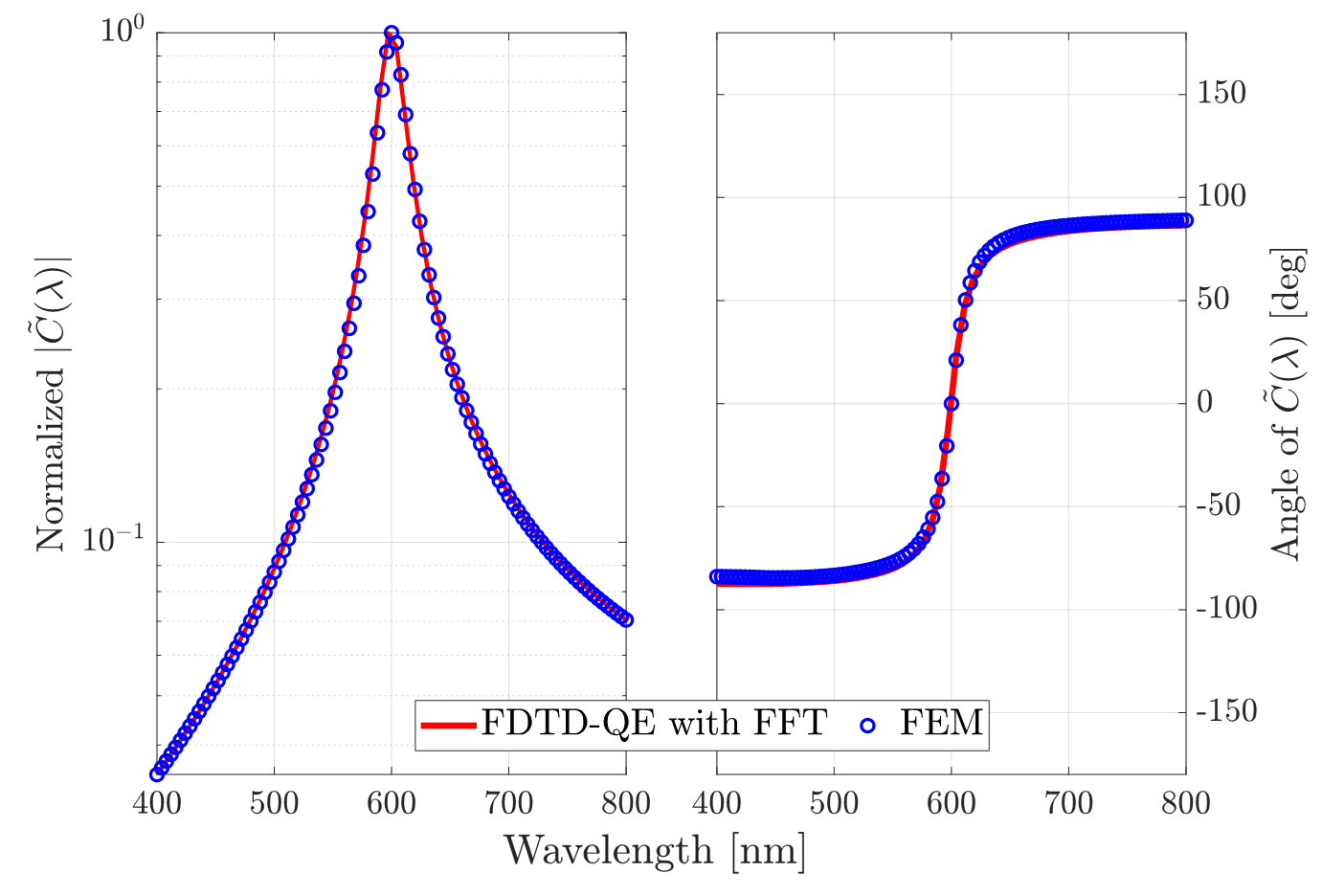}
        \caption{}
            \label{fig:FEM_cw}
    \end{subfigure}
    \begin{subfigure}[b]{\linewidth}
        \centering
        \includegraphics[width=\linewidth]{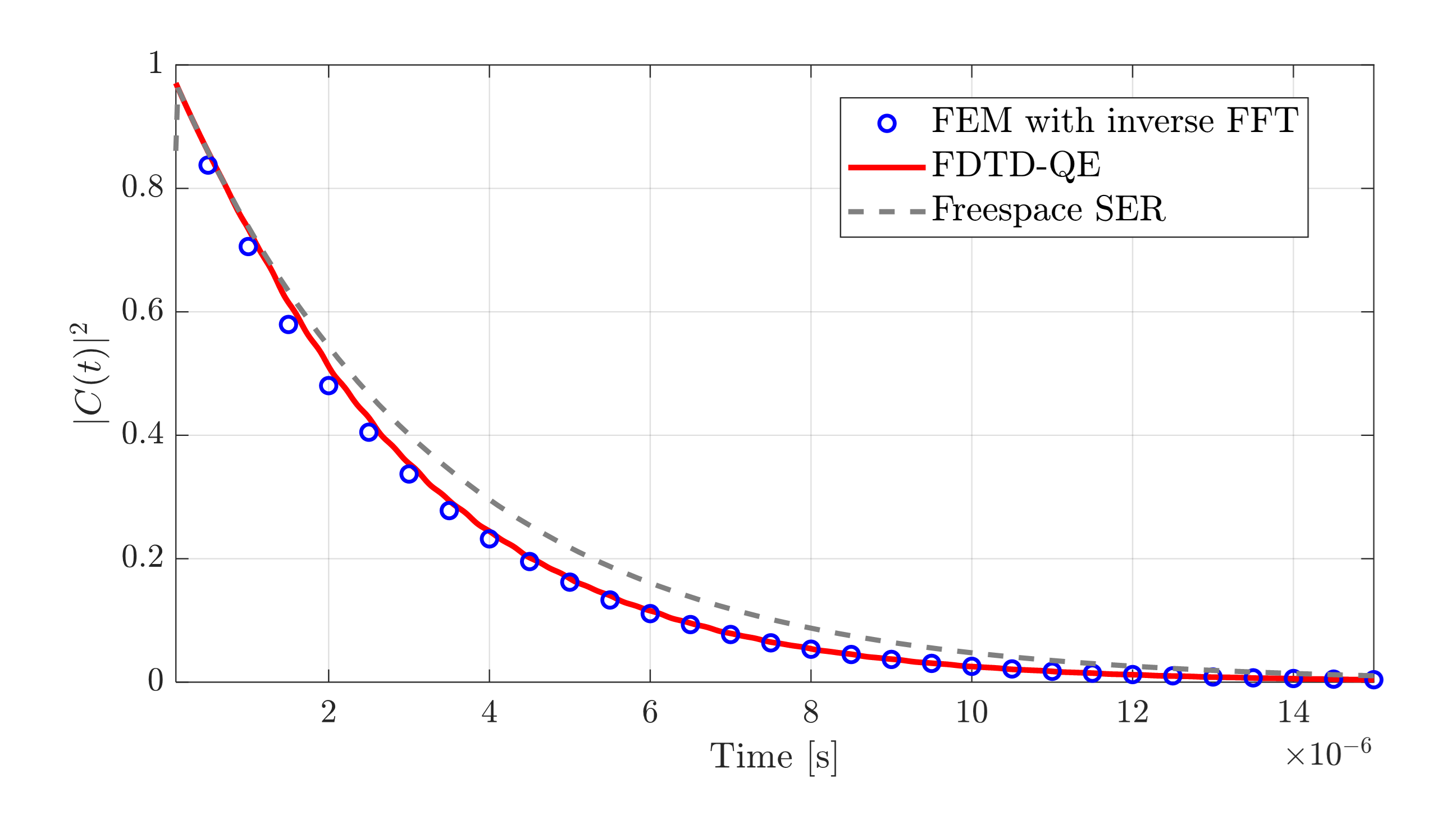}
        \caption{}
        \label{fig:FEM_ct}
    \end{subfigure}

    \caption{Verification of the proposed FEM by comparison with the FDTD-QE algorithm. (a) magnitude (left), angle (right) of $C(\omega)$ and (b) atomic population.}
    \label{fig:FEM_results}
\end{figure}

Next, we investigate the same physical configuration using the FEM, corresponding to the case previously analyzed with FDTD at $h = 482~\mathrm{nm}$. In the FEM method, the main focus is on evaluating the spectral response of the Green’s function. These spectra are used to calculate the atomic population in the frequency domain, as given in Eq.~\eqref{eq:C_omega}. As done for the calculation of the SFA in Fig.~\ref{fig:results}, we compute the imaginary part of the Green’s function spectrum over the wavelength range from 400 nm to 800 nm in this case. Using these Green’s functions, we transform them into the memory kernel in the frequency domain according to Eq.~\eqref{eq:Ka_def}, and subsequently obtain the spectrum of $C(\omega)$ from Eq.~\eqref{eq:C_omega}. To ensure a consistent comparison, the calculated $C(\omega)$ was compared with the fast Fourier transform (FFT) spectrum of the time-domain result obtained from the FDTD-QE simulation, as presented in Fig.~\ref{fig:FEM_cw}.
By applying an inverse FFT to 
$C(\omega)$, we recover the time-domain amplitude $C(t)$, from which the atomic population $|C(t)|^2$ is obtained. The reconstructed result is then compared with the direct time-domain result obtained from the FDTD-QE simulation as shown in Fig.~\ref{fig:FEM_ct}.
In both cases, the results show good agreement with the FDTD-QE simulations, demonstrating that the proposed approach accurately reproduces the emission dynamics of the metallic half-space structure.

\subsection{TLS centered in planar Fabry-Perot cavity}

We next construct a symmetric configuration by placing an identical metallic mirror on the opposite side of the atom, thereby forming a Fabry--Perot cavity to study its emission dynamics. 
Such a configuration supports multiple reflections between the two mirrors, leading to strong interference effects that significantly modify the local density of optical states (LDOS) and, consequently, the spontaneous emission dynamics.

\begin{figure}[b]
    \centering
    \includegraphics[width=\linewidth]{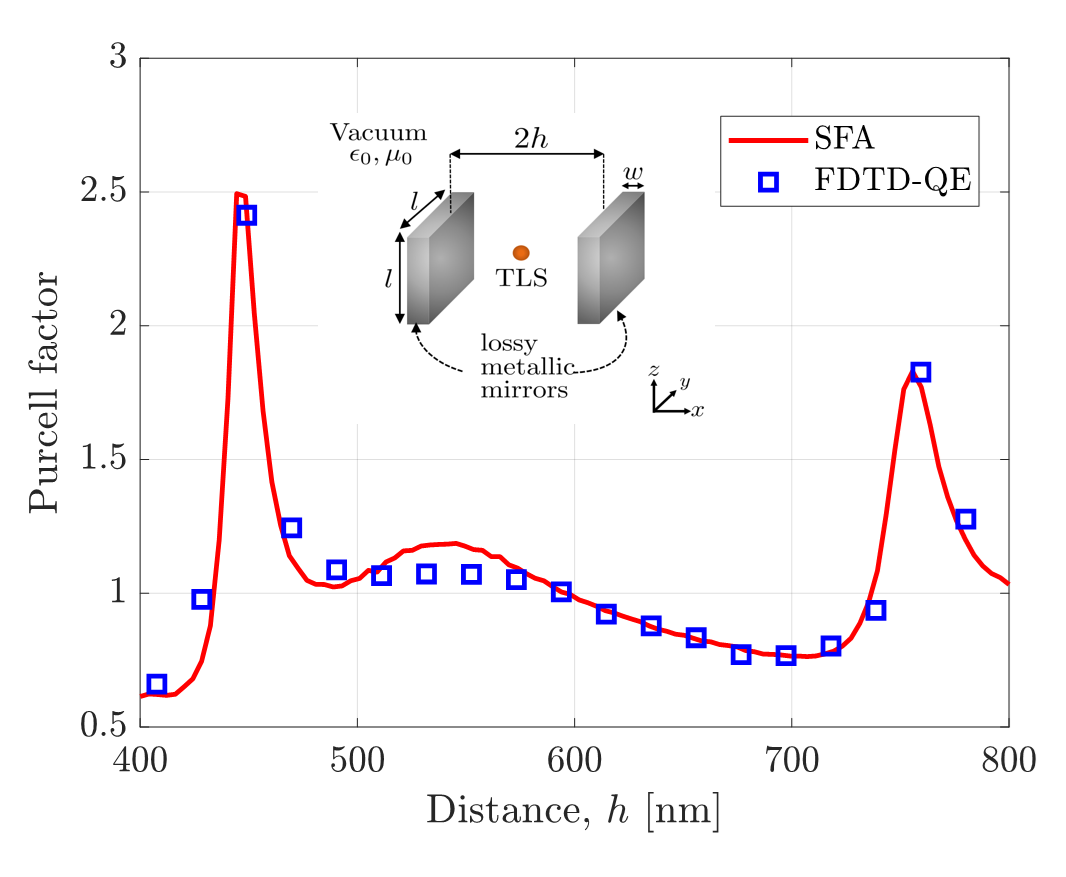}
    \caption{Purcell factor as a function of the emitter–surface distance $h$ for a two-level system (TLS) centered in a metallic cavity structure.}
    \label{fig:Two_mirror_distance}
\end{figure}

Except for the reduced dipole moment of $d_{a} = 1\times10^{-8}~\mathrm{C{\cdot}m}$, 
all other parameters remain the same as in the previous single-mirror case. 
Accordingly, the separation between the two cavity mirrors is $2h$, 
with the TLS always located at the center of the cavity. 
As in the previous case, we vary the mirror-to-TLS distance $h$ to investigate the change in the Purcell factor as a function of the cavity spacing, 
as illustrated in Fig.~\ref{fig:Two_mirror_distance}.
The results clearly show that, unlike the moderately oscillating behavior observed in the single-mirror case in Fig.~\ref{fig:purcell_distance}, 
the two-mirror cavity exhibits a pronounced enhancement of the Purcell factor at specific cavity lengths corresponding to the Fabry--Perot resonances. 

\begin{figure}
    \centering
    \subfloat[]{%
        \includegraphics[width=0.48\linewidth]{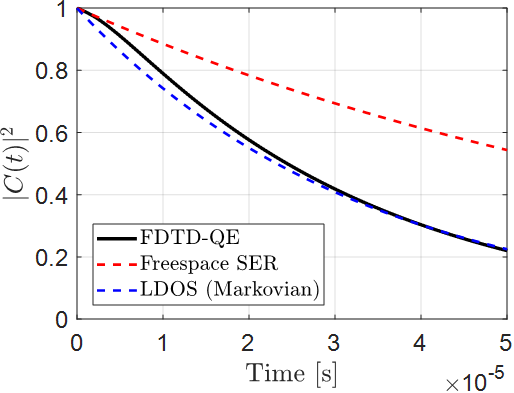}%
        \label{fig:pop_fast2} 
    }
    \subfloat[]{%
        \includegraphics[width=0.48\linewidth]{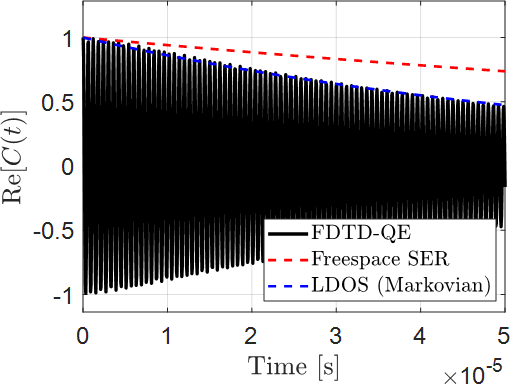}%
        \label{fig:pop_slow2}
    }
    \caption{Time dynamics at the cavity length corresponding to the maximum Purcell factor ($h=453$ nm): (a) atomic population decay and (b) $\text{Re}[C(t)]$.}
                        
    \label{fig:dynamics_all2}
\end{figure}

This demonstrates that strong field confinement and constructive interference between the mirrors lead to a significant increase in the spontaneous emission rate at those resonant separations.
We also investigate the atomic population dynamics at the cavity length where this resonance-induced Purcell enhancement reaches its maximum ($h=453$ nm), as depicted in Fig.~\ref{fig:dynamics_all2}. At this separation, the emitter exhibits a pronounced acceleration of spontaneous decay, and the result clearly captures the non-Markovian features within the resonant cavity, in contrast to the smooth exponential decay governed by the Markovian LDOS in free space.
Although a slight population revival associated with vacuum Rabi oscillations emerges at later times, only the dynamics prior to this revival are shown here, since the cavity quality factor $Q$ and atom–field coupling strength are relatively modest in this case.
\subsection{TLS centered in concave Fabry-Perot cavity}
To improve field confinement and enhance light–matter interaction, we consider a cylindrically curved concave Fabry–Perot cavity operating at the third longitudinal mode with a cavity length $L$ of $1.5\lambda_a$ ($h = 450~\mathrm{nm}$), a mirror curvature radius $R = L$, and an aperture angle $\alpha = 52^\circ$. The material parameters are identical to those used in the previous examples, and TLS is placed at the center of the cavity. The parameters related to the TLS are the same as those in Section~V-A.

\begin{figure}
    \centering
    \includegraphics[width=\linewidth]{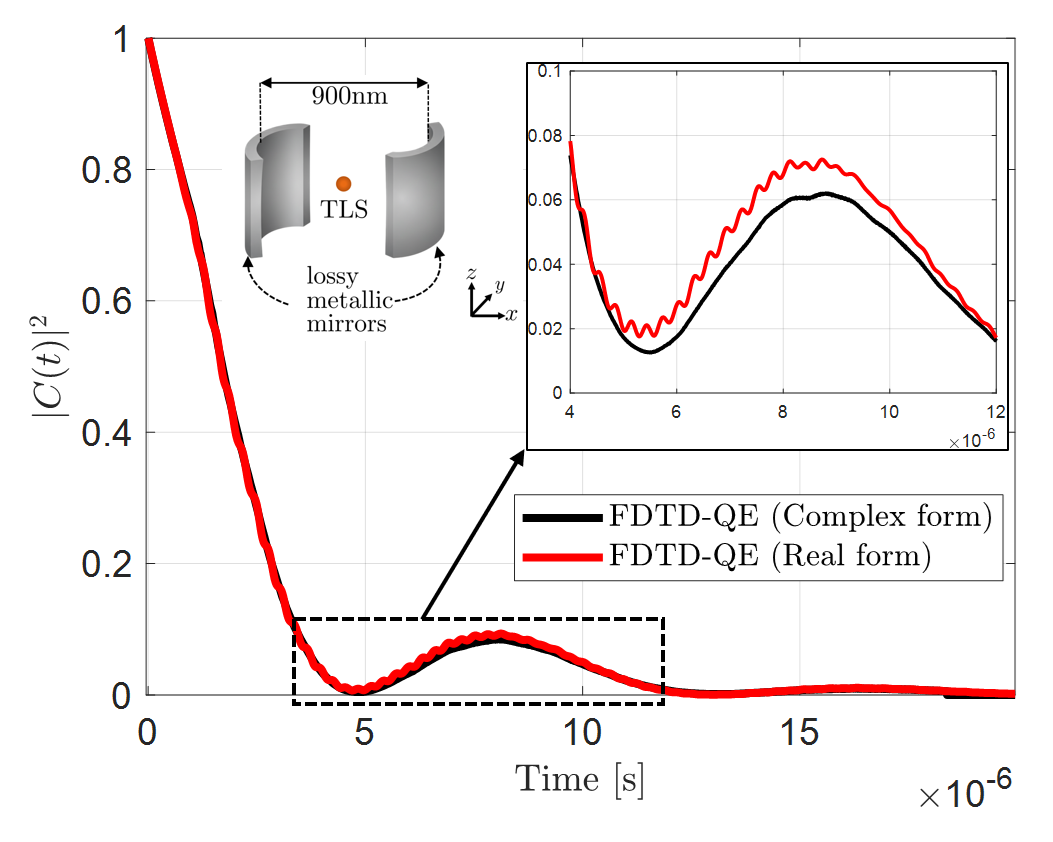}
    \caption{Comparison of the population dynamics $|C(t)|^2$ obtained from the complex and real-valued (approximate) formulations.}
    \label{fig:Weak_rabi_Ct2}
\end{figure}

In contrast to the previous planar cavity, the concave geometry introduces moderate radiative feedback, producing a small but distinct modulation in the population—a clear signature of the onset of vacuum Rabi–type oscillations.
The corresponding result of $C(t)$ is shown in Fig.~\ref{fig:Weak_rabi_Ct2}. While the cavity geometry is intentionally simplified for numerical verification and has not been optimized to achieve maximal field confinement or ideal Rabi oscillations, it nevertheless provides a clear and quantitative demonstration of coherent atom–field exchange under realistic dissipative conditions.
Note that, unlike conventional approaches where material losses are introduced phenomenologically through damping constants or empirical decay terms, the present framework derives the interaction self-consistently from first principles by incorporating the full Lorentz–Drude dispersion into the time-domain field equations.
This treatment enables direct modeling of radiative feedback and energy dissipation, accurately capturing the non-Markovian dynamics of the emitter–field interaction.

The population obtained from the complex and real-valued (approximate) formulations are nearly identical under modest atom–field coupling. However, as the coupling strength increases, the approximation in Eq.~\eqref{eq:J_tls_final} gradually deviates from the exact complex formulation in Eq.~\eqref{eq:J_tls_time_append}. As shown in Fig.~\ref{fig:Weak_rabi_Ct2}, the real-valued implementation exhibits unphysical high-frequency oscillations associated with $\omega_a$, which are absent in the complex formulation. As the oscillation period approaches the radiative lifetime $T_{\mathrm{rad}}$, or equivalently when the coupling enters a stronger regime, these fluctuations could become more pronounced, resulting in a noticeable deviation between the two results.

\begin{figure}
    \centering
    \subfloat[]{%
        \includegraphics[width=0.48\linewidth]{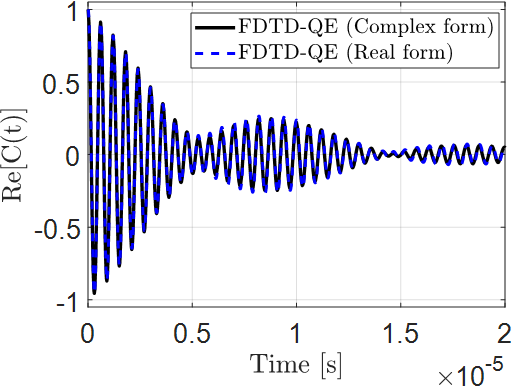}%
        \label{fig:Rabi_ReC} 
    }
    \hfill
    \subfloat[]{%
        \includegraphics[width=0.48\linewidth]{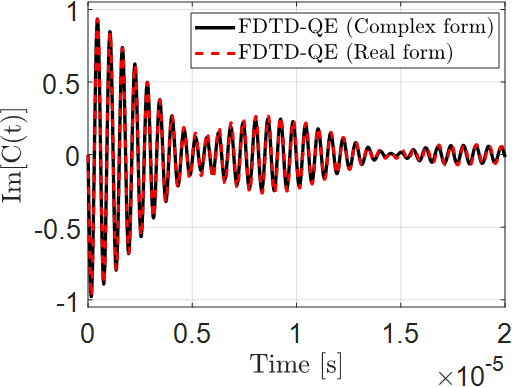}%
        \label{fig:Rabi_ImC}
    }

    \caption{Temporal dynamics of the real (a) and imaginary (b) components of the emitter amplitude $C(t)$ obtained from the complex and real-valued (approximate) formulations.}
    \label{fig:dynamics_all}
\end{figure}

The real and imaginary components of $C(t)$, shown in Figs.~\ref{fig:Rabi_ReC} - ~\ref{fig:Rabi_ImC}, oscillate at the optical transition frequency $\omega_a$, and the slowly varying envelope clearly exhibits the revival behavior characteristic of vacuum Rabi oscillations.

\begin{figure}
    \centering
    \includegraphics[width=\linewidth]{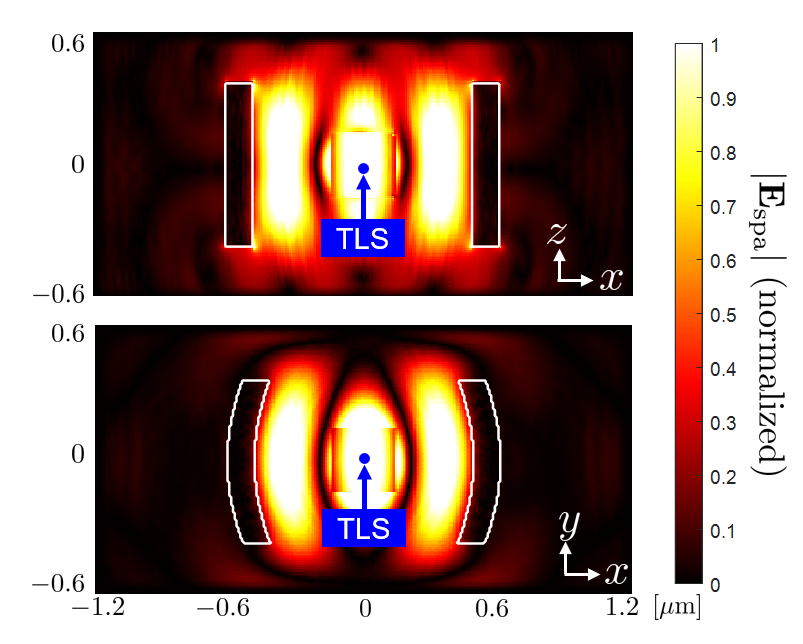}
    \caption{Spatial distribution of the normalized single-photon amplitude in the $x$–$z$ and $x$–$y$ planes, where the concave cavity boundaries are outlined in white.}
    \label{fig:SPA_field_2}
\end{figure}

Fig.~\ref{fig:SPA_field_2} shows the normalized magnitude of single-photon amplitude in the $x$–$z$ and $x$–$y$ planes, with the concave cavity boundaries outlined in white. A clear standing-wave pattern with three intensity maxima appears along the cavity axis, confirming operation at the third longitudinal mode with a cavity length of about $1.5\lambda_a$. The confined field near the center and the alternating node–antinode structure indicate strong radiative feedback between the mirrors, consistent with the population modulation and vacuum Rabi oscillations described above.
Although this field is modeled through an equivalent current in the CEM framework, it represents the single-photon amplitude emitted from the TLS, fundamentally distinct from a classical electromagnetic field.


\section{conclusion}
In this work, we have developed a rigorous computational framework for modeling single-photon emission from two-level quantum systems in open and dissipative electromagnetic environments, going beyond the conventional Markovian approximation. Central to our approach is the modified Langevin noise formalism, which introduces boundary-assisted (BA) and medium-assisted (MA) field modes as a complete basis. By numerically validating the BA–MA mode completeness in fully three-dimensional settings, we established the fundamental link between the dyadic Green’s function and quantum emitter dynamics.

We also numerically verified that the discrete BA–MA expansion accurately reconstructs the imaginary part of the Green’s function in both free-space and PEC-confined scenarios. This validation provides strong evidence that atomic population dynamics and single-photon amplitudes can be equivalently and efficiently obtained via the Green’s function, without the need for explicit computation of all BA–MA modes. Building on this foundation, we demonstrated practical implementations within both FDTD and FEM solvers, incorporating Lorentz–Drude materials and showing consistency with analytical predictions.

Overall, this framework extends classical computational electromagnetics into the quantum regime, enabling the first-principles modeling of non-Markovian light–matter interactions. Beyond verifying fundamental theory, it also offers a practical toolset for simulating quantum emitters in realistic nanophotonic and plasmonic environments, paving the way for optimized single-photon sources, cavity quantum electrodynamics devices, and integrated quantum photonic platforms.

Future directions include extending the approach to multi-emitter systems, strong-coupling regimes, and time-dependent materials to explore richer classes of quantum electrodynamic phenomena. In particular, analyzing interaction effects and quantum correlations among multiple atoms within the CEM framework represents a promising avenue, as it would enable the study of collective emission, superradiance and entanglement dynamics in complex nanophotonic structures.

\bibliography{sorsamp}


\ifCLASSOPTIONcaptionsoff
  \newpage
\fi

\end{document}